\documentclass[twocolumn]{aa}
\usepackage{graphicx}
\usepackage{epsfig,graphicx,rotating,portland,fancyheadings,longtable,natbib}
\usepackage{txfonts}

\def\sax{{\it Beppo}SAX$~$}
\def\aro{{$\alpha_{\rm ro}$~}}
\def\aox{{$\alpha_{\rm ox}$~}}

\def\ergs{{erg~cm$^{-2}$s$^{-1}$~}}
\def\ergj{{erg~cm$^{-2}$s$^{-1}$Jy$^{-1}$~}}
\newcommand{\lsim}{{\lower.5ex\hbox{$\; \buildrel < \over \sim \;$}}}
\newcommand{\gsim}{{\lower.5ex\hbox{$\; \buildrel > \over \sim \;$}}}

\def\fxfr{$f_{\rm x}/f_{\rm r}$~}
\def\nupeak{$\nu_{peak}$~}

\begin{document}
\title{The Sedentary Survey of Extreme High Energy Peaked BL Lacs. \\ 
II. The Catalog and Spectral Properties}
\author{ P.~Giommi\inst{1}
\and S.~Piranomonte\inst{1}
\and M.~Perri\inst{1}
\and P.~Padovani\inst{2}$^,$\inst{3}$^,$\inst{4}
\institute{
ASI Science Data Center, ASDC, Agenzia Spaziale Italiana c/o ESRIN, via G. Galilei 00044 Frascati, Italy
\and
Space Telescope Science Institute, 3700 San Martin Drive, Baltimore, MD 21218, USA
\and
Affiliated to the Space Telescope Division of the European Space Agency, ESTEC, Noordwijk, the Netherlands
\and
ST-ECF, European Southern Observatory, Karl-Schwarzschild-Str. 2, D-85748 Garching bei M\"unchen, Germany 
(current address)
}}
\offprints{paolo.giommi@asi.it}
\date{Received ....; accepted ....}

\authorrunning{P. Giommi et al.}
	\titlerunning{The Sedentary Survey of Extreme High Energy Peaked BL Lacs. II.}


\abstract{
The multi-frequency `Sedentary Survey' is a deep, statistically complete, radio flux 
limited sample comprising 150 BL Lacertae objects distinguished by their extremely 
high X-ray to radio flux ratio (\fxfr), ranging from five hundred to over five 
thousand times that of typical BL Lacs discovered in radio surveys. 
This large excess of high energy photons compared to radio emission is thought to be 
due to synchrotron radiation that in these sources reaches the UV or the X-ray band. 
The name `Sedentary Survey` originates from the multi-frequency technique used
to select the sample that was expected to be so efficient as to allow the 
conduction of some preliminary statistical studies even without the need to identify the 
candidates through optical spectroscopy. 
The details of the selection criteria and the preliminary results have been 
published in \cite{Gio99}. In this paper we present the final, 100\% identified,
catalog together with the optical, X-ray and broad-band Spectral Energy 
Distributions (SED) constructed combining literature multi-frequency data with 
non-simultaneous optical observations and \sax X-ray data, when available.
The SEDs confirm that the peak of the synchrotron power in these objects is located 
at very high energies. \sax wide band X-ray observations show that, in most cases, 
the X-ray spectra are convex and well described by a logarithmic parabola model 
peaking (in a $\nu f(\nu)~vs~\nu$ representation) between 0.02 to several keV. 
Although detailed X-ray spectral data are available for only about one fifth of the sources 
the observed peaks never reach energies well above 10 keV (as in Mkn 501 
during the large X-ray flare of April 1997 and in 1ES 2344+514 in December 1996) implying 
that hard X-ray synchrotron peak energies are rare and probably associated with strong 
flaring events.

Owing to the high synchrotron energies involved most of the sources in the catalog 
are likely to be TeV emitters, with the closest and brightest ones probably detectable 
by the present generation of Cherenkov telescopes. However, only 50\% (3 out of 6) of the 
presently established TeV BL Lacs are actually included in the survey suggesting that the 
hardest peaks 
may be associated with secondary synchrotron components that can be detected only above the 
soft X-ray band. The existence of secondary emission regions is suggested by the 
strong X-ray spectral curvature that in some objects predicts an optical flux much 
below the observed emission.

The optical spectrum of about one fourth of the sources is totally featureless 
hampering any red-shift or luminosity determination. Because this implies that 
the non-thermal nuclear emission must be well above that of the host galaxy, these objects 
are likely to be the most powerful sources in the survey and therefore be examples of the yet 
unreported {\it high radio luminosity--high energy peaked} BL Lacs. 
The existence of such objects would be at odds with the claimed inverse proportionality 
between radio power and synchrotron peak energy known as the ``blazar sequence''. 

At the low-power end of the luminosity dynamical range, where the non-thermal optical continuum 
falls below the emission from the host galaxy, recognition issues start becoming important since 
BL Lacs in this luminosity regime can hardly be recognized as such, but rather as {\it radio galaxies} 
or simply as {\it elliptical galaxies}. We have found a small sample of bright nearby 
elliptical galaxies that are candidate low radio power high energy peaked BL Lacs. 


}

\maketitle

\section{Introduction}

BL Lacertae objects are a rare and very peculiar type of Active Galactic Nuclei  
(AGN). Their observational properties, which include super-luminal motion, 
strong and rapidly variable non-thermal radiation across the entire electromagnetic  
spectrum and a high degree of polarization, are believed to be the signature of strongly 
amplified radiation emitted in a relativistic jet closely aligned to the line of sight 
(e.g. \cite{Urry95}).  

These unusual physical and geometrical properties, combined with the peculiar cosmological 
evolution that distinguishes BL Lacs from other types of AGN, have made this class of 
sources the subject of intense research activity and of large multi-frequency observation 
campaigns.

Despite the fact that BL Lacs emit strongly over the entire electromagnetic spectrum, 
nearly all of presently known sources of this type have been discovered at radio or at 
X-ray frequencies, or through a combination of these two bands. 
However, in some still poorly explored observing windows, like the millimeter/microwave 
region and the gamma-ray/TeV bands, BL Lacs, together with Flat Spectrum Radio Quasars (FSRQ), 
are expected to be one of the major constituents of the extra-galactic discrete 
source population. New large samples of these objects will certainly be built when deep surveys 
based on data from the Planck and GLAST space missions will become available in a few years.

The observed non-thermal emission in BL Lacs is thought to be due to synchrotron 
emission peaking (in a $Log(\nu f(\nu)-Log(\nu)$ representation) between the 
far infrared and the hard X-ray band, followed by Inverse Compton scattering  
up to very high energies. Those BL Lacs where the synchrotron peak is located at low 
energy (known as Low energy peaked BL Lacs, LBL, \cite{PG95}) so far have been 
discovered mostly in radio  surveys, while those where the synchrotron power reaches 
the UV or the X-ray band  (High energy peaked BL Lacs, or HBL) have been discovered 
much more frequently in X-ray surveys. 

The Sedentary Multi-frequency Survey (\cite{Gio99}, hereafter  
referred to as Paper I) was designed to assemble a large and statistically well  
defined sample of HBL BL Lacs by exploiting the fact that the electromagnetic 
emission of these sources is so extreme that no other type of extra-galactic source  
type is known to possess a similar Spectral Energy Distribution (SED).
By imposing radio, optical and X-ray flux ratios that are only 
consistent with the unique SEDs of HBL BL Lacs it is then possible to  
build large samples of these rare objects with very high selection efficiency.   

The sample presented in Paper I included 155 BL Lac candidates, only 40\% of 
which were at the time spectroscopically identified. However, 
it was estimated that the multi-frequency selection technique applied ensured 
that at least 85\% of the candidates were genuine BL Lacs. 
That allowed the authors to derive, though in a preliminary way, some important 
statistical properties of HBL BL Lacs, such as their radio LogN-LogS and  
Cosmological evolution. 
For that reason the survey was named ``Sedentary''. 

The estimation of some of the fundamental properties of the sample, however, require 
the knowledge of luminosity, hence redshift which makes an identification campaign 
clearly necessary. This need prompted the organization of a dedicated optical 
spectroscopy observation program (see \cite{paperIII}, hereafter Paper III) that, 
together with data collected by other independent groups, mostly aimed at 
the systematic identification of bright high Galactic latitude RASS sources 
(\cite{Schwope00}, \cite{Bauer00}, \cite{beck00}, \cite{anderson03}), led to the identification of {\it all} 
the candidates in the sample.

In this paper we present the complete cleaned sample, which now includes 150 objects following
i) the removal of those candidates that the spectroscopic identification campaign did not confirm
to be BL Lacs, and
ii) the addition of 7 new BL Lacs that satisfy all criteria for inclusion in the survey but
were not in the original sample because their \aro was just below the threshold of 0.2 due to
the optical contamination from the host galaxy which was not taken into account.

We also present a detailed spectral analysis based on broad band \sax archival data and the 
radio to X-ray SED of a selection of objects built using multi-frequency literature data, on our own
optical observations and \sax data, when available.

The radio LogN-LogS, luminosity function and cosmological evolution have been presented in preliminary
form in Paper I and in \cite{Perri02}, the final results are presented in a dedicated paper 
(\cite{paperIV}, hereafter Paper IV). 

\section{The sample}

A complete description of the Sedentary Survey sample is given in Paper I,
in this section we summarize the main selection criteria and we refer the reader 
to the original paper for more details.  
 
The sample was extracted from a large set of radio and X-ray emitting sources 
selected through a cross-correlation between the RASS catalog of bright X-ray  
sources (\cite{Vog99}) and the NVSS catalog of radio (1.4 GHz, \cite{Con98}) sources. 
The following conditions were imposed to avoid the complications due to the Galactic 
plane and ensure that the sample is statistically complete above the radio flux limit 
of $f_{\rm r} = 3.5$ mJy 
 
\begin{enumerate} 
\item $|b|>20^{\circ}$; 
\item \fxfr $\ge 3\times 10^{-10}$ \ergj; 
\item $\alpha_{\rm ro} > 0.2$; 
\item $f_{\rm r} \ge 3.5$ mJy; 
\item RASSBSC count rate $\ge 0.1$ cts/s; 
\item $V \le 21$; 
\end{enumerate}  
 
\noindent
where $\alpha_{\rm ro}$ is the usual broad band spectral index between 
the radio (5GHz) and optical (5000 $\AA$) fluxes and $V$ is the visual  
apparent magnitude of the optical counterpart.  
 
Condition 1) limits the survey area to high Galactic latitude regions where  
soft X-ray absorption due to Galactic $N_H$ is low;  
condition 2) imposes a very large \fxfr flux ratio that can be reached by HBL BL Lacs only; 
condition  3) removes from the sample radio quiet sources, such as nearby Seyfert  
galaxies where the unrelated radio and X-ray flux may accidentally  
satisfy condition 2); condition 4), 5) and 6) are necessary to ensure  
statistical completeness above $f_{\rm r} \ge 3.5$ mJy. 

\begin{figure}[ht]
\vbox{
\centerline{
\includegraphics[width=7cm] {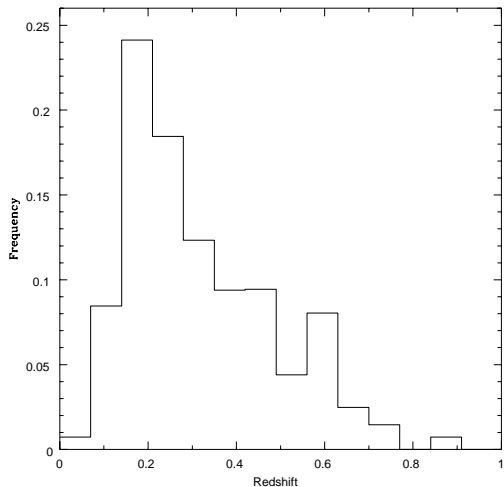} 
}}
\caption{The redshift distribution of the subsample of 111 BL Lacs with a measured redshift after 
de-convolving for the X-ray sky coverage of the survey (see also \cite{L01}).
}
\label{z_dist}
\end{figure}

\section {The Catalog}

The fully identified complete sample including 150 extreme HBL BL Lacs is presented 
in Table 1 where column 1 gives the source name built with the catalog identification code
SHBL (where S stands for ``Sedentary'' survey and HBL for High energy peaked BL Lacs, 
\cite{PG95}) and the arc-second precision optical coordinates of the source taken from 
the APM (\cite{Irw94}) and COSMOS (\cite{Yen92}) on-line services; column 2 gives the RASS name; 
column 3, 4 and 5 give the X-ray flux (0.1-2.4 keV), the 
radio flux (20 cm, from the NVSS survey), and the optical apparent $V$ magnitude (from APM and COSMOS, 
see Paper I) respectively; column 6 gives the redshift, when available; column 7 gives 
the reference for the optical identification. 


The catalog is also available on the web at the following address

\begin{center} 
http://www.asdc.asi.it/sedentary/ 
\end{center} 

\noindent
where additional data, including the broad band spectral energy distributions, finding 
charts and optical spectra (from Paper III) are also provided, when available. 

Redshifts were measured for 111 BL Lacs, mostly from spectral features due to 
the host galaxy. The other 39 sources ($\sim 25\%$ of the sample) remain without 
redshift since their optical spectrum does not show any emission or absorption lines. 

\begin{figure}[ht]
\vbox{
\centerline{
\includegraphics[width=7cm] {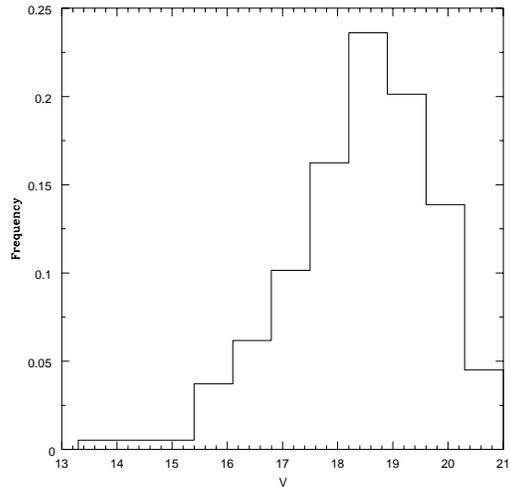}
}}
\caption{The sky coverage corrected distribution of optical apparent magnitudes ($V$) of all sources in the Sedentary 
survey. Note that the fraction of sources with $V$ $\ge 20$ is very low suggesting that the condition
6 in the definition of the sample ($V$ $\le 21$) only excluded a tiny fraction of sources. 
}
\label{V_dist}
\end{figure}

The sample is flux limited, complete (that is {\it all} the sources above the flux limit are included) 
and 100\% identified and is therefore suitable for an unbiased investigation 
of the statistical properties of the population that it represents. 

The main cosmological properties, 
such as radio LogN-LogS, luminosity function and cosmological evolution, are studied in detail in Paper 
IV; in the following we provide only some basic statistics of the parameters listed in Table 1. 
The redshift and $V$ distributions, corrected for the X-ray sky coverage of the survey, 
are shown in Figs.~\ref{z_dist} and \ref{V_dist} respectively. 
The average redshift of the subsample of 111 objects for which a value could be 
measured is $\langle z \rangle = 0.32$ and is rather low compared to other 
radio loud AGN of similar or even higher radio flux (e.g. \cite{P03}), but consistent with 
earlier results about X-ray selected BL Lacs that have demonstrated the peculiar cosmological 
evolution of these sources (\cite{Bade98}, \cite{Rec00}). 
However, the exclusion of the 39 sources without a redshift determination, which may  well
be distant, high luminosity objects (see below) implies that this value is likely to be only
a lower limit.   

The distribution of optical magnitudes (see Fig.~\ref{V_dist}) is 
sharply peaked around the mean value of $\langle V \rangle = 18.4$ 
and only a few sources are fainter than  $V \approx 20$.
Condition number 6 in the sample definition criteria (i.e. $V \le 21$) should therefore 
exclude a very small fraction of BL Lacs ($ \lsim 1-2\%$, see also Paper 1).

\subsection {Notes on individual objects} 
 
\subsubsection{SHBL J040128.0+815312} This source has been identified as an  
early-type galaxy by \cite{Bauer00} but its X-ray and radio luminosities are rather  
high (about $5\times10^{44} \rm{erg/s}$ and $3\times10^{31} \rm{erg/s/Hz}$ respectively).  
It is therefore likely that this source is a AGN, in the following we will  
assume that it is a BL Lac.             

\subsubsection{SHBL~J114535.1$-$034001, SHBL~J235023.2$-$243603}
These X-ray/radio sources are within clusters of galaxies.  
However, since i) the RASS X-ray emission in these sources is not extended, ii) the radio emission 
coincides with a galaxy, and iii) from our optical spectroscopic campaign we have found that the 
Ca H\&K break in their optical spectrum is diluted by non-thermal radiation. We assume that these 
objects are BL Lacs in clusters.

\onecolumn
\setcounter{table}{0}
\begin{table}[*t]
\caption{The complete sample of HBL BL Lacs in the Sedentary Survey: objects coordinates and main properties} 
\begin{center}
\begin{tabular}{llcrlcc} 
\hline 
\multicolumn{1}{l}{Source name}&\multicolumn{1}{c}{RASS name}  &f$_{0.1-2.4~\mathrm{keV}}$ & f$_{20~\mathrm{cm}}$ & \multicolumn{1}{c}{$V$}& \multicolumn{1}{c}{$z$} & \multicolumn{1}{c}{Ref. for }\\ 
&\multicolumn{1}{c}{(1RXS J)}&  [${\rm erg/cm^{2}/s}$] & [mJy] & & & opt. ID\\ 
\multicolumn{1}{c}{(1)}&\multicolumn{1}{c}{(2)} & \multicolumn{1}{c}{(3)} & \multicolumn{1}{c}{(4)} & (5) & (6) & (7) \\ 
\hline 
SHBL J001355.9$-$185406& 001356.6$-$18540& 1.26$\times 10^{-11}$& 29.6& 16.8  &  0.095 & (a) \\         
SHBL J001527.9$+$353639& 001528.3$+$35364& 3.45$\times 10^{-12}$& 11.2& 18.3  &  ...     & ($^*$) \\       
SHBL J001827.8$+$294729& 001827.8$+$29473& 1.43$\times 10^{-11}$& 33.8& 18.5  &  0.100 & (b)  \\         
SHBL J003334.2$-$192133& 003334.6$-$19213& 1.49$\times 10^{-11}$& 18.9& 16.1  &  0.610 & (a), ($^ *$) \\ 
SHBL J003514.7$+$151504& 003514.9$+$15151& 7.33$\times 10^{-12}$& 18.8& 17.1  &  0.250 & (a)  \\         
SHBL J004208.0$+$364112& 004208.1$+$36411& 3.68$\times 10^{-12}$& 12.0& 18.0  &  ...     &(a), ($^*$) \\   
SHBL J005816.8$+$172310& 005817.1$+$17230& 8.38$\times 10^{-12}$& 9.4 & 19.2  &  ...     & (a)  \\         
SHBL J011050.0$-$125502& 011050.0$-$12545& 1.51$\times 10^{-11}$& 17.5& 17.4  &  0.234 & (b)  \\         
SHBL J011501.9$-$340028& 011501.3$-$34000& 2.20$\times 10^{-12}$& 6.4 & 20.1  &  0.482 & ($^*$) \\       
SHBL J011747.0$-$244333& 011746.6$-$24432& 3.82$\times 10^{-12}$& 10.3& 19.0  &  0.279 & ($^*$) \\       
SHBL J012308.7$+$342049& 012308.9$+$34204& 6.14$\times 10^{-11}$& 45.7& 15.2  &  0.272 & (c)  \\         
SHBL J012338.2$-$231058& 012338.2$-$23110& 9.03$\times 10^{-12}$& 27.6& 18.7  &  0.404 & (a)  \\         
SHBL J012657.2$+$330730& 012657.1$+$33073& 4.75$\times 10^{-12}$& 7.1 & 17.5  &  ...     & ($^*$) \\       
SHBL J013632.5$+$390559& 013632.9$+$39055& 2.33$\times 10^{-11}$& 60.6& 15.4  &  ...     &(a), ($^*$) \\   
SHBL J014040.7$-$075848& 014040.9$-$07585& 1.19$\times 10^{-11}$& 27.4& 18.3  &  ...     & (d)  \\         
SHBL J020106.3$+$003401& 020106.3$+$00340& 6.85$\times 10^{-12}$& 13.4& 18.0  &  0.299 & (e)  \\         
SHBL J020412.8$-$333342& 020413.6$-$33334& 2.54$\times 10^{-12}$& 6.4 & 18.6  &  0.617 & ($^*$) \\       
SHBL J020838.1$+$352313& 020837.5$+$35231& 6.99$\times 10^{-12}$& 5.0 & 19.2  &  0.318 & (f)  \\         
SHBL J021630.9$+$231513& 021632.3$+$23144& 1.40$\times 10^{-11}$& 35.9& 19.2  &  0.288 & (g)  \\         
SHBL J022716.6$+$020158& 022716.6$+$02015& 3.93$\times 10^{-11}$& 37.0& 18.2  &  ...     & (a)  \\         
SHBL J023536.6$-$293843& 023536.7$-$29384& 2.81$\times 10^{-12}$& 5.5 & 17.6  &  ...     & (a)  \\         
SHBL J025018.8$-$212940& 025018.2$-$21295& 4.78$\times 10^{-12}$& 3.8 & 19.8  &  0.498 & (a)  \\         
SHBL J030330.2$+$055430& 030330.0$+$05542& 1.16$\times 10^{-11}$& 29.7& 17.4  &  0.196 & (b)  \\         
SHBL J030416.3$-$283217& 030416.4$-$28321& 4.95$\times 10^{-12}$& 8.3 & 19.3  &  ...     & (d)  \\         
SHBL J031422.9$+$061956& 031422.7$+$06200& 2.29$\times 10^{-11}$& 29.4& 18.0  &  ...     & (a)  \\         
SHBL J031633.7$-$221611& 031634.6$-$22161& 2.44$\times 10^{-12}$& 4.4 & 19.0  &  0.228 & ($^*$) \\       
SHBL J031951.9$+$184534& 031951.9$+$18453& 2.70$\times 10^{-11}$& 22.9& 18.1  &  0.190 & (f)  \\         
SHBL J032350.7$+$071736& 032350.5$+$07174& 6.69$\times 10^{-12}$& 4.5 & 20.3  &  ...     & ($^*$) \\       
SHBL J032541.0$-$164618& 032540.8$-$16460& 5.87$\times 10^{-11}$& 27.6& 16.0  &  0.291 & (a)  \\         
SHBL J032613.9$+$022515& 032613.6$+$02252& 3.36$\times 10^{-11}$& 68.3& 17.4  &  0.147 & (h)  \\         
SHBL J034923.2$-$115927& 034922.8$-$11592& 3.09$\times 10^{-11}$& 24.9& 18.2  &  0.185 & (i)  \\         
SHBL J035856.1$-$305448& 035855.6$-$30543& 4.86$\times 10^{-12}$& 13.2& 18.5  &  ...     & (d)  \\         
SHBL J040128.0$+$815312& 040129.1$+$81532& 3.55$\times 10^{-12}$& 9.7 & 17.6  &  0.215 & (a)  \\         
SHBL J040324.5$-$242950& 040324.1$-$24293& 5.01$\times 10^{-12}$& 7.4 & 20.2  &  0.357 & ($^*$) \\       
SHBL J041112.2$-$394143& 041112.1$-$39413& 4.10$\times 10^{-12}$& 5.3 & 18.8  &  ...     & ($^*$) \\       
SHBL J041652.4$+$010524& 041652.6$+$01053& 7.32$\times 10^{-11}$& 120.7&17.5  &  0.287 & (h)  \\         
SHBL J042132.8$-$062905& 042132.7$-$06284& 6.45$\times 10^{-12}$& 16.0& 20.1  &  0.390 & (a)  \\         
SHBL J042218.4$+$195051& 042218.2$+$19504& 9.98$\times 10^{-12}$& 9.0 & 20.3  &  0.516 & (f)  \\         
SHBL J042900.1$-$323640& 042900.5$-$32363& 5.81$\times 10^{-12}$& 9.3 & 17.6  &  ...     & (d)  \\         
SHBL J044018.5$-$245933& 044017.8$-$24591& 4.06$\times 10^{-12}$& 12.7& 19.5  &  ...     & (d)  \\         
SHBL J044127.4$+$150456& 044127.8$+$15045& 3.93$\times 10^{-11}$& 14.0& 19.8  &  0.109 & ($^*$) \\       
SHBL J044230.1$-$001830& 044229.8$-$00182& 4.06$\times 10^{-12}$& 4.2 & 20.0  &  0.449 & ($^*$) \\       
SHBL J050335.3$-$111507& 050335.6$-$11150& 1.29$\times 10^{-11}$& 10.5& 17.9  &  ...     & ($^*$) \\       
SHBL J050939.0$-$040036& 050938.3$-$04003& 2.70$\times 10^{-11}$& 71.2& 19.0  &  0.304 & (h)  \\         
SHBL J050939.8$-$251403& 050940.0$-$25135& 2.13$\times 10^{-12}$& 4.5 & 20.1  &  0.264 & ($^*$)  \\        
SHBL J060714.3$-$251859& 060714.2$-$25185& 4.13$\times 10^{-12}$& 12.2& 18.9  &  0.275 & ($^*$) \\         
SHBL J062149.6$-$341149& 062150.0$-$34114& 4.47$\times 10^{-12}$& 8.8 & 18.7  &  0.529 & ($^*$) \\         
SHBL J071219.0$+$571948& 071218.9$+$57193& 3.13$\times 10^{-12}$& 8.0 & 19.9  &  0.095 & (l)  \\           
SHBL J074405.6$+$743358& 074405.6$+$74335& 1.36$\times 10^{-11}$& 23.4& 16.9  &  0.315 & (f)  \\           
SHBL J075124.9$+$173051& 075124.3$+$17304& 3.85$\times 10^{-12}$& 10.5& 17.4  &  0.185 & (a), ($^*$) \\    
SHBL J075324.6$+$292132& 075322.4$+$29215& 2.78$\times 10^{-12}$& 4.3 & 18.7  &  0.161 & ($^*$) \\         
SHBL J083251.3$+$330009& 083251.9$+$33001& 3.33$\times 10^{-12}$& 4.5 & 20.8  &  0.671 & (m)  \\           
SHBL J084712.9$+$113350& 084713.3$+$11334& 2.38$\times 10^{-11}$& 33.3& 17.2  &  0.199 & (n)  \\           
SHBL J085909.9$+$834459& 085916.5$+$83445& 3.63$\times 10^{-12}$& 10.3& 17.5  &  0.327 & (o)  \\           
SHBL J091322.2$+$813304& 091324.6$+$81331& 3.56$\times 10^{-12}$& 5.1 & 19.8  &  0.639 & (o)  \\           
\hline 
\end{tabular} 
\end{center}
\end{table} 
 
\setcounter{table}{0}
\begin{table}[*t]
\caption{The complete sample of HBL BL Lacs in the Sedentary Survey: objects coordinates and main properties - Continued} 
\begin{center}
\begin{tabular}{llcrlcc} 
\hline 
\multicolumn{1}{l}{Source name}&\multicolumn{1}{c}{RASS name}  &f$_{0.1-2.4~\mathrm{keV}}$ & f$_{20~\mathrm{cm}}$ & \multicolumn{1}{c}{$V$}& \multicolumn{1}{c}{$z$} & \multicolumn{1}{c}{Ref. for }\\ 
&\multicolumn{1}{c}{(1RXS J)}&  [${\rm erg/cm^{2}/s}$] & [mJy] & & & opt. ID\\ 
\multicolumn{1}{c}{(1)}&\multicolumn{1}{c}{(2)} & \multicolumn{1}{c}{(3)} & \multicolumn{1}{c}{(4)} & (5) & (6) & (7) \\ 
\hline 
SHBL J092401.1$+$053345& 092401.1$+$05335& 5.15$\times 10^{-12}$& 7.6  & 18.4  & ...     & (a), ($^*$) \\    
SHBL J093037.5$+$495025& 093037.1$+$49502& 2.69$\times 10^{-11}$& 21.6 & 17.2  & 0.188 & (c) \\            
SHBL J094355.5$-$070951& 094355.3$-$07094& 2.88$\times 10^{-12}$& 7.3  & 19.9  & 0.433 & ($^*$) \\         
SHBL J095224.0$+$750213& 095225.8$+$75021& 4.83$\times 10^{-12}$& 12.4 & 17.2  & 0.179 & (a), ($^*$) \\    
SHBL J095805.9$-$031740& 095806.4$-$03172& 2.82$\times 10^{-12}$& 7.6  & 19.6  & ...     & ($^*$) \\         
SHBL J100656.4$+$345445& 100656.9$+$34544& 2.05$\times 10^{-12}$& 6.7  & 19.0  & 0.612 & (o)  \\           
SHBL J100811.4$+$470519& 100811.5$+$47052& 1.13$\times 10^{-11}$& 4.9  & 18.4  & 0.343 & (m)  \\           
SHBL J101015.9$-$311908& 101015.9$-$31190& 2.83$\times 10^{-11}$& 74.4 & 17.3  & 0.143 & ($^*$)    \\      
SHBL J101616.7$+$410812& 101616.4$+$41081& 7.59$\times 10^{-12}$& 15.0 & 18.9  & 0.270 & (p)     \\        
SHBL J102212.6$+$512358& 102212.5$+$51240& 5.16$\times 10^{-12}$& 5.3  & 18.1  & 0.141 & (f)     \\        
SHBL J102243.8$-$011302& 102244.2$-$01125& 1.31$\times 10^{-11}$& 36.3 & 15.5  & ...   & (a), ($^*$) \\      
SHBL J103118.6$+$505335& 103118.6$+$50534& 5.58$\times 10^{-11}$& 38.1 & 16.6  & 0.361 & (q)     \\        
SHBL J104651.4$-$253545& 104651.9$-$25354& 4.53$\times 10^{-12}$& 14.3 & 19.2  & 0.254 & ($^*$)    \\      
SHBL J105125.3$+$394325& 105125.1$+$39432& 4.48$\times 10^{-12}$& 11.0 & 18.4  & 0.498 & (o)     \\        
SHBL J105606.6$+$025213& 105607.0$+$02521& 1.49$\times 10^{-11}$& 4.5  & 17.7  & 0.236 & (a)    \\         
SHBL J105723.1$+$230317& 105723.5$+$23031& 6.41$\times 10^{-12}$& 8.1  & 18.7  & 0.379 & (r)    \\         
SHBL J110021.1$+$401927& 110021.3$+$40193& 7.60$\times 10^{-12}$& 18.4 & 17.8  & 0.225 & (o)    \\         
SHBL J110337.6$-$232931& 110337.7$-$23293& 5.09$\times 10^{-11}$& 121.1& 16.1  & 0.186 & (h)    \\         
SHBL J110427.3$+$381231& 110427.1$+$38123& 2.71$\times 10^{-10}$& 768.6& 14.4  & 0.031 & (s)     \\        
SHBL J111130.9$+$345203& 111131.2$+$34521& 7.05$\times 10^{-12}$& 8.6  & 19.2  & 0.212 & (m)     \\        
SHBL J111706.3$+$201407& 111706.3$+$20141& 5.42$\times 10^{-11}$& 103.2& 15.7  & 0.139 & (g)     \\        
SHBL J111939.5$-$304720& 111941.0$-$30465& 3.72$\times 10^{-12}$& 9.6  & 19.5  & 0.412 & ($^*$)    \\      
SHBL J112048.0$+$421212& 112047.5$+$42121& 1.37$\times 10^{-11}$& 24.2 & 17.3  & 0.124 & (c)    \\         
SHBL J112348.9$+$722958& 112349.2$+$72300& 6.98$\times 10^{-12}$& 12.6 & 18.5  & ...     & (g)     \\        
SHBL J113444.5$-$172902& 113443.6$-$17285& 3.64$\times 10^{-12}$& 5.1  & 19.7  & 0.571 & ($^*$)    \\      
SHBL J113630.3$+$673704& 113630.9$+$67370& 2.38$\times 10^{-11}$& 45.9 & 16.7  & 0.135 & (t)       \\      
SHBL J113755.4$-$171042& 113755.4$-$17103& 4.72$\times 10^{-12}$& 5.3  & 18.9  & 0.600 & ($^*$)    \\      
SHBL J114535.1$-$034001& 114535.8$-$03394& 7.96$\times 10^{-12}$& 19.8 & 18.4  & 0.167 & (d), ($^*$) \\    
SHBL J114755.0$+$220539& 114754.9$+$22054& 3.77$\times 10^{-12}$& 4.3  & 18.3  & 0.276 & (o)      \\        
SHBL J114930.3$+$243926& 114930.4$+$24392& 9.73$\times 10^{-12}$& 28.6 & 18.0  & 0.402 & (o)       \\         
SHBL J115404.4$-$001010& 115404.9$-$00100& 4.81$\times 10^{-12}$& 10.7 & 18.7  & 0.253 & (e)        \\        
SHBL J121158.6$+$224232& 121158.1$+$22423& 8.25$\times 10^{-12}$& 20.3 & 18.5  & 0.455 & (o) \\            
SHBL J122121.9$+$301037& 122121.7$+$30104& 3.16$\times 10^{-11}$& 71.6 & 16.4  & 0.182 & (a)      \\         
SHBL J123417.1$-$385635& 123416.9$-$38563& 6.10$\times 10^{-12}$& 7.0  & 18.0  & 0.236 & ($^*$)    \\      
SHBL J123511.0$-$140322& 123511.1$-$14033& 2.63$\times 10^{-12}$& 4.2  & 19.8  & 0.407 & ($^*$)    \\      
SHBL J123705.5$+$302004& 123705.6$+$30200& 1.06$\times 10^{-11}$& 5.8  & 20.5  & 0.700 & (m)            \\   
SHBL J123739.0$+$625842& 123739.2$+$62584& 4.03$\times 10^{-12}$& 12.6 & 18.6  & 0.297 & (f)            \\   
SHBL J124149.3$-$145558& 124149.8$-$14555& 1.81$\times 10^{-11}$& 17.3 & 17.3  & ... & (c), ($^*$) \\         
SHBL J125015.5$+$315559& 125015.0$+$31560& 1.88$\times 10^{-12}$& 5.7  & 19.5  & ...     & ($^*$)    \\      
SHBL J125134.9$-$295843& 125135.2$-$29582& 4.37$\times 10^{-12}$& 10.5 & 19.0  & 0.487 & (c)            \\    
SHBL J125300.9$+$382626& 125301.0$+$38262& 6.45$\times 10^{-12}$& 4.9  & 19.2  & 0.372 & (a)            \\    
SHBL J125341.1$-$393159& 125341.2$-$39320& 1.96$\times 10^{-11}$& 50.1 & 18.3  & 0.179 & ($^*$)    \\       
SHBL J125731.9$+$241240& 125731.7$+$24124& 1.17$\times 10^{-11}$& 14.9 & 15.4  & 0.141 & (c)            \\    
SHBL J125847.8$-$044745& 125847.7$-$04474& 7.93$\times 10^{-12}$& 4.3  & 18.9  & 0.586 & (a)            \\    
SHBL J131155.7$+$085342& 131156.0$+$08534& 4.74$\times 10^{-12}$& 5.5  & 19.6  & 0.469 & (a)            \\    
SHBL J133529.7$-$295037& 133530.6$-$29503& 7.24$\times 10^{-12}$& 10.8 & 19.1  & 0.513 & (f)            \\    
SHBL J140630.1$-$393509& 140630.3$-$39350& 4.04$\times 10^{-12}$& 8.8  & 19.8  & ...     & ($^*$)    \\       
SHBL J140630.2$+$123620& 140630.1$+$12363& 2.56$\times 10^{-12}$& 6.3  & 20.6  & ...     & ($^*$)    \\       
SHBL J140659.2$+$164207& 140659.1$+$16420& 5.43$\times 10^{-12}$& 8.4  & 18.0  & ...     &(a), ($^*$)    \\   
SHBL J140918.9$+$135239& 140918.9$+$13525& 2.87$\times 10^{-12}$& 7.3  & 20.4  & 0.580 & (a)            \\      
SHBL J141756.1$+$254356& 141756.8$+$25432& 2.70$\times 10^{-11}$& 89.7 & 16.0  & 0.237 & (h)            \\       
SHBL J142239.0$+$580155& 142239.1$+$58015& 2.50$\times 10^{-11}$& 13.4 & 18.4  & 0.638 & (m)            \\      
SHBL J142739.5$-$252102& 142740.6$-$25210& 4.70$\times 10^{-12}$& 3.7  & 18.9  & 0.318 & ($^*$)    \\       
SHBL J142832.6$+$424024& 142832.6$+$42402& 5.25$\times 10^{-11}$& 58.9 & 16.4  & 0.130 & (h)            \\      
SHBL J143917.4$+$393243& 143917.7$+$39324& 1.79$\times 10^{-11}$& 42.9 & 16.6  & 0.344 & (a), ($^*$)    \\  
\hline 
\end{tabular} 
\end{center}
\end{table} 
 
\setcounter{table}{0}
\begin{table}[*t] 
\caption{The complete sample of HBL BL Lacs in the Sedentary Survey: objects coordinates and main properties - Continued} 
\begin{center}
\begin{tabular}{llcrlcc} 
\hline 
\multicolumn{1}{l}{Source name}&\multicolumn{1}{c}{RASS name}  &f$_{0.1-2.4~\mathrm{keV}}$ & f$_{20~\mathrm{cm}}$ & \multicolumn{1}{c}{$V$}& \multicolumn{1}{c}{$z$} & \multicolumn{1}{c}{Ref. for }\\ 
&\multicolumn{1}{c}{(1RXS J)}&  [${\rm erg/cm^{2}/s}$] & [mJy] & & & opt. ID\\ 
\multicolumn{1}{c}{(1)}&\multicolumn{1}{c}{(2)} & \multicolumn{1}{c}{(3)} & \multicolumn{1}{c}{(4)} & (5) & (6) & (7) \\ 
\hline 
SHBL J144506.3$-$032612& 144505.9$-$03261& 7.80$\times 10^{-12}$& 21.8  & 17.4 & ...     &(a), ($^*$)    \\   
SHBL J150340.6$-$154113& 150343.0$-$15410& 2.39$\times 10^{-11}$& 5.9   & 17.5 & ...     &(a), ($^*$)\\       
SHBL J150637.0$-$054004& 150636.4$-$05401& 5.03$\times 10^{-12}$& 15.3  & 19.5 & 0.518 & ($^*$)    \\       
SHBL J151041.0$+$333504& 151040.8$+$33351& 4.45$\times 10^{-12}$& 9.1   & 17.0 & 0.112 & (d), ($^*$)    \\  
SHBL J151618.6$-$152343& 151618.7$-$15234& 1.46$\times 10^{-11}$& 8.6   & 18.7 & ...     &(a), ($^*$)    \\   
SHBL J151747.4$+$652523& 151747.3$+$65252&  2.30$\times 10^{-11}$& 37.8 & 15.9 & 0.702 & (u)            \\  
SHBL J153311.3$+$185428& 153311.7$+$18542&  1.42$\times 10^{-11}$& 23.0 & 17.7 & 0.305 & (a), ($^*$)    \\  
SHBL J153500.9$+$532037& 153501.1$+$53204&  1.79$\times 10^{-11}$& 18.4 & 17.6 & 0.890 & (m)            \\  
SHBL J160518.9$+$542059& 160518.5$+$54210&  6.26$\times 10^{-12}$& 7.8  & 19.0 & 0.21  & (e)            \\  
SHBL J161204.6$-$043815& 161204.4$-$04381&  4.40$\times 10^{-12}$& 4.3  & 18.9 & ...     & ($^*$)    \\       
SHBL J161632.9$+$375602& 161633.4$+$37555&  1.47$\times 10^{-12}$& 4.6  & 18.7 & 0.204 & ($^*$)    \\       
SHBL J163123.5$+$421703& 163124.7$+$42165&  6.59$\times 10^{-12}$& 7.5  & 19.2 & 0.468 & (m)            \\  
SHBL J163658.4$-$124837& 163658.7$-$12483&  8.74$\times 10^{-12}$& 26.3 & 20.1 & 0.246 & ($^*$)    \\       
SHBL J174702.3$+$493801& 174702.0$+$49380&  2.81$\times 10^{-12}$& 7.9  & 19.9 & 0.460 & ($^*$)    \\       
SHBL J175615.9$+$552217& 175615.5$+$55221&  1.48$\times 10^{-11}$& 16.9 & 17.6 & ...     & (a), ($^*$)    \\   
SHBL J175713.4$+$703336& 175712.8$+$70333&  8.68$\times 10^{-12}$& 10.9 & 18.3 & 0.407 & (f)            \\  
SHBL J184822.3$+$653657& 184822.6$+$65370&  4.29$\times 10^{-12}$& 9.7  & 18.3 & 0.364 & ($^*$)    \\       
SHBL J203844.8$-$263633& 203845.1$-$26362&  4.73$\times 10^{-12}$& 5.7  & 18.5 & 0.437 & ($^*$)    \\       
SHBL J204735.8$-$290859& 204737.0$-$29090&  3.97$\times 10^{-12}$& 10.7 & 19.3 & 0.333 & ($^*$)    \\       
SHBL J204921.7$+$003926& 204921.6$-$00393&  4.85$\times 10^{-12}$& 6.0  & 18.1 & 0.256 & ($^*$)    \\       
SHBL J205242.7$+$081038& 205242.6$+$08103&  4.90$\times 10^{-12}$& 6.2  & 19.6 & ...     & ($^*$)    \\       
SHBL J213135.4$-$091523& 213135.5$-$09152&  1.58$\times 10^{-11}$& 43.6 & 16.7 & 0.449 & (a), ($^*$)    \\  
SHBL J213151.3$-$251558& 213151.7$-$25160&  6.04$\times 10^{-12}$& 11.0 & 17.3 & ...     & ($^*$)    \\      
SHBL J213852.5$-$205348& 213852.9$-$20535&  1.33$\times 10^{-11}$& 11.5 & 17.9 & 0.290 & (a), ($^*$)    \\ 
SHBL J215852.0$-$301331& 215852.2$-$30133&  5.72$\times 10^{-10}$& 490.3& 13.5 & 0.117 & (v)            \\ 
SHBL J220155.8$-$170702& 220156.0$-$17065&  5.12$\times 10^{-12}$& 4.8  & 18.1 & 0.169 & (a)            \\ 
SHBL J222253.8$-$175321& 222253.9$-$17531&  2.88$\times 10^{-12}$& 5.7  & 19.4 & 0.297 & ($^*$)    \\      
SHBL J223812.7$-$394020& 223812.7$-$39401&  6.21$\times 10^{-12}$& 20.4 & 18.6 & 0.250 & (d)            \\ 
SHBL J224340.1$-$123100& 224341.9$-$12310&  9.45$\times 10^{-12}$& 10.7 & 18.1 & 0.226 & (b)            \\ 
SHBL J224910.7$-$130002& 224911.1$-$13000&  9.73$\times 10^{-12}$& 7.6  & 18.9 & ...     & (a), ($^*$)    \\ 
SHBL J225147.3$-$320614& 225146.9$-$32061&  3.61$\times 10^{-12}$& 3.6  & 19.0 & ...     & (a), ($^*$)    \\ 
SHBL J230436.8$+$370507& 230437.1$+$37050&  1.85$\times 10^{-11}$& 23.1 & 17.8 & ...     & (a), ($^*$)    \\ 
SHBL J230634.9$-$110348& 230636.0$-$11035&  5.19$\times 10^{-12}$& 11.5 & 19.2 & ...     & (a)            \\ 
SHBL J230722.0$-$120518& 230722.5$-$12052&  2.94$\times 10^{-12}$& 7.3  & 18.6 & ...     & ($^*$)    \\      
SHBL J230846.7$-$221949& 230846.7$-$22195&  8.29$\times 10^{-12}$& 6.5  & 16.0 & 0.137 &(f)            \\  
SHBL J231028.0$-$371909& 231027.0$-$37192&  1.97$\times 10^{-12}$& 6.3  & 17.8 & ...     & ($^*$)    \\      
SHBL J234333.8$+$344004& 234332.5$+$34395&  1.51$\times 10^{-11}$& 35.0 & 20.1 & 0.366 & ($^*$)    \\      
SHBL J235023.2$-$243603& 235023.6$-$24355&  2.14$\times 10^{-12}$& 6.8  & 18.3 & 0.193 & (a), ($^*$)    \\ 
SHBL J235730.0$-$171805& 235730.1$-$17180&  2.14$\times 10^{-11}$& 44.5 & 17.3 & ...     &(d)         \\     
SHBL J235907.9$-$303739& 235908.0$-$30374&  6.50$\times 10^{-11}$& 65.0 & 17.0 & 0.165 &(a)            \\  
\hline 
\end{tabular} 
\end{center}
 
\vspace{0.5cm} 
($^*$) \cite{paperIII};
(a) \cite{Bauer00};
(b) \cite{Fis98}; 
(c) \cite{PG95};
(d) \cite{Schwope00};
(e) \cite{Sloan01} ;
(f) \cite{Rec00};
(g) \cite{Bor00};
(h) \cite{NED92};
(i) \cite{Scha93};
(l) \cite{beck00};
(m) \cite{Bade98};
(n) \cite{L01};
(o) \cite{beck03};
(p) \cite{Cao99};
(q) \cite{Pol97};
(r) \cite{White00};
(s) \cite{deVau};
(t) \cite{Bade94};
(u) \cite{beck99};
(v) \cite{Fal93}.
\end{table}
\twocolumn

\begin{table*}[hb]
\begin{center}
\caption{List of candidate low luminosity HBLs} 
\begin{tabular}{lllcccc}
\hline
\multicolumn{1}{c}{Source name}  &\multicolumn{1}{c}{RASS name}&\multicolumn{1}{l}{Other name}& 
\multicolumn{1}{c}{Redshift}     &\multicolumn{1}{c}{$V$}  &\multicolumn{1}{c}{X-ray luminosity}&
\multicolumn{1}{c}{Classification} \\ 
& \multicolumn{1}{c}{(1RXS J)} & & & &(erg/s)&\\
\multicolumn{1}{c}{(1)} &\multicolumn{1}{c}{(2)} &\multicolumn{1}{c}{(3)}&(4) &(5) &(6) &(7) \\
\hline
SHBL J020014.8+312545 & 020014.5+31254&NGC0777 &0.017&12.8&$5.6\times10^{42}$& Elliptical \\
SHBL J025251.8$-$011629 & 025250.3$-$01160&NGC1132 &0.023&7.6&$1.1\times10^{43}$& Elliptical \\
SHBL J033851.7$-$353536 & 033851.5$-$35354&NGC1404 &0.006&10.5&$5.9\times10^{41}$& Elliptical \\
SHBL J044255.9$-$263509 & 044255.9$-$26345&1WGA J0442.9-2635 &N.A.&15.5& ... & S0 \\
SHBL J150111.2+014208 & 150111.6+01415&NGC5813 &0.006&11.7&$2.2\times10^{42}$& Elliptical \\
\hline
\end{tabular}
\end{center}
\end{table*}

\begin{table*}[h]
\label{rejected}
\begin{center}
\caption{List of the 19 emission line AGNs in the HBL zone}
\begin{tabular}{llcccccc}
\hline
\multicolumn{1}{c}{RASS name}& 
f$_{0.1-2.4~\mathrm{keV}}$ & f$_{20~\mathrm{cm}}$ & \multicolumn{1}{c}{$V$} & \multicolumn{1}{c}{$z$} 
& \multicolumn{1}{c}{Luminosity$_{20~\mathrm{cm}}$} & \multicolumn{1}{c}{\aro}&
\multicolumn{1}{c}{Ref. for} \\ 
& [${\rm erg/cm^{2}/s}$] & [mJy] & & &[${\rm erg/s/Hz}$] & & opt. ID\\
\multicolumn{1}{c}{(1)} & \multicolumn{1}{c}{(2)} & (3) & (4) & (5) & (6) & (7) & (8) \\
\hline

1RXS J000729.3$+$02405 & 3.08$\times 10^{-12}$ &  7.6 &   18.0  &  0.300 &1.68$\times 10^{31}$ &0.26 & (a)   \\
1RXS J013526.9$-$04263 & 9.04$\times 10^{-12}$ &  8.2 &   17.5  &  0.155 &4.32$\times 10^{30}$ &0.24 & (b)  \\
1RXS J023727.6$-$26302 & 2.52$\times 10^{-12}$ &  5.6 &   20.2  &  0.141 &2.41$\times 10^{30}$ &0.24 & ($^*$)   \\
1RXS J023832.1$+$02333 & 4.10$\times 10^{-12}$ & 10.4 &   17.1  &  0.209 &1.04$\times 10^{31}$ &0.22 & (c)   \\
1RXS J035018.4$-$22170 & 7.00$\times 10^{-12}$ & 19.1 &   16.6  &  0.111 &4.98$\times 10^{30}$ &0.24 & (d)   \\
1RXS J035245.6$-$23425 & 4.30$\times 10^{-12}$ &  4.3 &   18.1  &  0.140 &1.83$\times 10^{30}$ &0.23 & (e) \\
1RXS J035432.5$-$13400 & 4.90$\times 10^{-12}$ & 15.8 &   17.0  &  0.077 &1.92$\times 10^{30}$ &0.25 & (b)   \\
1RXS J084206.6$+$07593 & 8.33$\times 10^{-12}$ & 17.4 &   17.8  &  0.134 &6.73$\times 10^{31}$ &0.32 & (f)   \\
1RXS J092554.3$+$40041 & 4.56$\times 10^{-12}$ & 13.2 &   17.4  &  0.470 &8.09$\times 10^{31}$ &0.26 & (g)   \\
1RXS J093318.2$-$17144 & 6.25$\times 10^{-12}$ & 20.2 &   16.7  &  0.313 &4.91$\times 10^{31}$ &0.24 & (b)   \\
1RXS J122044.5$+$69053 & 5.38$\times 10^{-12}$ &  8.2 &   17.1  &  0.110 &2.10$\times 10^{30}$ &0.21 & (h)   \\
1RXS J123802.1$+$36164 & 1.89$\times 10^{-12}$ &  4.9 &   19.4  &  0.383 &1.84$\times 10^{31}$ &0.33 & ($^*$)   \\
1RXS J130350.5$-$39503 & 4.96$\times 10^{-12}$ & 13.6 &   17.4  &  0.121 &4.17$\times 10^{30}$ &0.27 & ($^*$)   \\
1RXS J133950.5$+$15593 & 2.01$\times 10^{-12}$ &  4.6 &   18.0  &  0.277 &8.45$\times 10^{30}$ &0.22 & ($^*$)   \\
1RXS J170817.7$-$03493 & 5.10$\times 10^{-12}$ &  4.8 &   17.5  &  0.180 &3.48$\times 10^{30}$ &0.19 & ($^*$)   \\
1RXS J172202.0$+$43152 & 6.27$\times 10^{-12}$ &  9.0 &   18.4  &  0.139 &3.76$\times 10^{30}$ &0.31 & (b) \\
1RXS J182042.7$+$38171 & 6.59$\times 10^{-12}$ &  7.5 &   18.8  &  0.077 &9.88$\times 10^{29}$ &0.33 & ($^*$)   \\
1RXS J212516.7$-$25553 & 2.72$\times 10^{-12}$ &  5.2 &   18.2  &  0.343 &1.54$\times 10^{31}$ &0.25 & ($^*$)   \\
1RXS J222944.5$-$27553 & 2.39$\times 10^{-12}$ &  7.8 &   18.8  &  0.322 &2.02$\times 10^{31}$ &0.33 & ($^*$)   \\    
\hline
\end{tabular}
\end{center}
\vspace{0.5cm} 

($^*$) our data;
(a) \cite{Hew95};
(b) \cite{Bauer00};
(c) \cite{Schn94};
(d) \cite{Rei96};
(e) \cite{Vog99};
(f) \cite{Bade98b};
(g) \cite{Buc98};
(h) \cite{Puch92};
\end{table*}
\twocolumn

\section{Optical Spectroscopy} 

In Paper I only about 40\% of the sources in the sample were confirmed BL Lacs. 
Although several other candidates were subsequently identified in a number of projects 
dedicated to the optical follow up observations of bright Rosat X-ray sources 
(\cite{Schwope00}, \cite{Bauer00}, \cite{beck00}), many objects would have remained unidentified
without a dedicated optical spectroscopy program.  
We have therefore carried out an extensive optical identification campaign using the 
Kitt Peak National Observatory 4m telescope, the ESO 3.6m telescope at 
La Silla and the TNG 3.6m telescope at La Palma, which allowed us to identify 
{\it all} sources of the survey. The results are described in detail in Paper III; in the 
following we summarize the main results.

\begin{figure*}[ht]
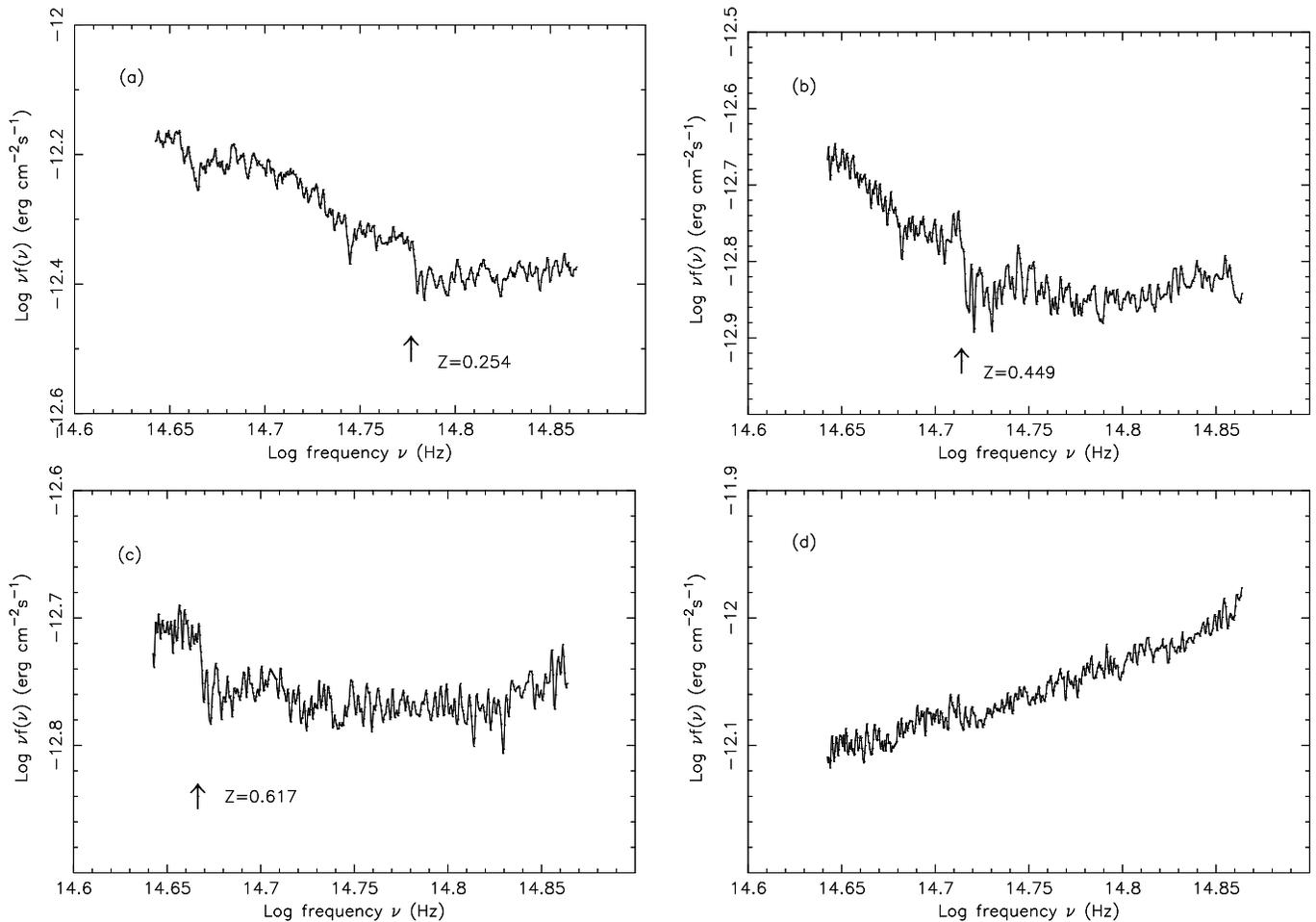

%
%
  \includegraphics[width=6.3cm,angle=-90]{sed_J1046m2535_opt.ps}
  \includegraphics[width=6.3cm,angle=-90]{sed_J0442m0018_opt.ps}
  \includegraphics[width=6.3cm,angle=-90]{sed_J0204m3333_opt.ps}
  \hspace{0.5cm}
  \includegraphics[width=6.3cm,angle=-90]{sed_J1503m1541_opt.ps}
  \caption{SED of the optical region of the four BL Lac objects SHBLJ~104651.4$-$253545 (a),  
           SHBLJ~044230.1$-$001830 (b), SHBLJ~020412.8$-$333342 (c) and
           SHBLJ~150340.6$-$154113 (d). Note that the optical output from the host galaxy (steep component) 
           dominates the spectra of the sources (a), (b) and (c) at frequencies below the Ca H\&K break 
           (indicated by the arrow) while the nuclear non-thermal emission (flat component) is apparent at 
           frequencies above the Ca H\&K break. The optical spectrum of source (d) is instead totally 
           dominated by its flat non-thermal component; no spectral features can be seen and therefore 
           its redshift cannot be determined.
  }
\label{opt_sed}
\end{figure*}

Good quality optical spectra were obtained for the 76 objects which were either previously 
unidentified (58) or had been reported in the literature as BL Lacs (18) but no redshift 
was given and no information about the quality of the optical spectrum was reported (see Table 1).   
Out of the 58 unclassified candidates 50 were confirmed to be BL Lacs and 
8 sources turned out to be emission line AGN. 
 
The full set of spectra are reported and discussed in Paper III. In this paper we show the 
optical spectra of four representative sources located at different redshifts, (see Fig.~\ref{opt_sed}) 
plotted in $Log(\nu f(\nu))~vs~Log(\nu) $ space which is normally used to plot broad band Spectral Energy 
Distributions (SED). We have chosen this graphical representation because it is 
particularly effective in enhancing broad features and changes in the spectral shape.

The data have been de-reddened and converted to units suitable for  
$\nu f(\nu)$ vs $\nu$ SED plots with the IRAF\footnote{IRAF is distributed by the 
National Optical Astronomy Observatories, which are operated by the Association of the 
Universities for Research in Astronomy, Inc., under cooperative agreement with the 
National Science Foundation} packages {\it noao.onedspec.deredden} and 
{\it noao.onedspec.splot}.

Figure \ref{opt_sed} illustrates how the emitted power at optical frequencies is a 
blend between the optical output from the host galaxy (steep component), which dominates 
the spectrum at frequencies below the Ca H\&K break, and the non-thermal nuclear emission 
(flat component), that usually appears at frequencies above the Ca H\&K break. 
The clear trend in the balance between the non-thermal and the galaxian component with 
redshift strongly indicates that high redshift (which on average implies high luminosity 
in a flux limited sample) sources are characterized by a flat, featureless optical spectrum.
In sources located at redshifts higher than 0.7-0.8 the Ca H\&K break falls outside the optical 
band and the non-thermal emission, which at these distances and radio flux above the 
survey limit is much brighter than the typical host galaxy, completely dominates the optical 
spectrum. 
This calls for a high redshift location of the 39 sources in the sample whose optical 
spectrum is flat and completely featureless (see discussion in Sect.~\ref{high_lum}). 
An estimate of a lower limit to the redshift of these sources will be given in Paper III and PaperIV.

\subsection{Elliptical galaxies or low luminosity BL Lacs?} 

\begin{figure}[ht]
\centerline{
\vbox{
\includegraphics[width=6.0cm,angle=-90]{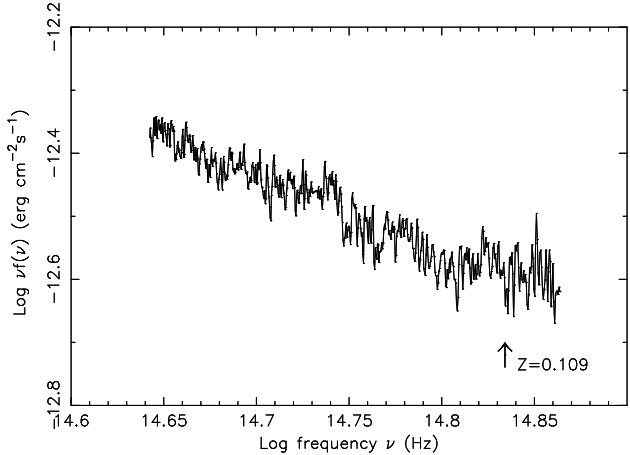}
}}
\caption{SED of the optical region of SHBLJ~044127.4+150456. Most of the emission from this nearby 
(z=0.109) object is red i.e. steep and dominated by the optical output from the host galaxy.
The nuclear non-thermal emission (flat component) is barely apparent at frequencies above the 
Ca H\&K break (indicated by the arrow).}
\label{opt_sed_1}
\end{figure}

Since the luminosity of BL Lac host galaxies (giant ellipticals) is approximately 
constant (\cite{Wur97}, \cite{Urry00}) the blend between the non-thermal nuclear emission and 
the optical flux from the host galaxy must be a strong function of the luminosity of the BL Lac.
This implies that at the low and high luminosity ends of the radio luminosity function the 
appearance of the optical spectrum of BL Lacs must be very different, as discussed in the
previous paragraph and shown in Fig.~\ref{opt_sed} (see also \cite{L02}). 
In particular, the optical emission of low redshift-low luminosity BL Lacs, 
must be almost completely dominated by the emission from the host galaxy. 
An example of this 
effect is shown in Fig.~\ref{opt_sed_1} where the non-thermal component of SHBL J044127.4+150456, 
a z=0.109 BL Lac, is barely detectable only at frequencies above the Ca H\&K break. 
At even lower redshifts (z$\lsim 0.1$) the host galaxy totally dominates the spectrum and distinguishing 
between normal radio galaxies and BL Lacs becomes very difficult.

In this luminosity regime the X-ray to radio flux ratio (\fxfr) remains unaffected, but
\aro is heavily contaminated by starlight and could decrease significantly 
pushing these low luminosity BL Lacs over the borderline between the HBL zone and the 
radio quiet zone (\aro $< 0.2$) thus biasing the sample at low luminosities. 
One object of this type, the elliptical galaxy IC1459, was recently associated with a BL Lac 
by \cite{Giommi02}.

\begin{figure}[ht]
\vbox{
\centerline{
\includegraphics[width=8.0cm,angle=0] {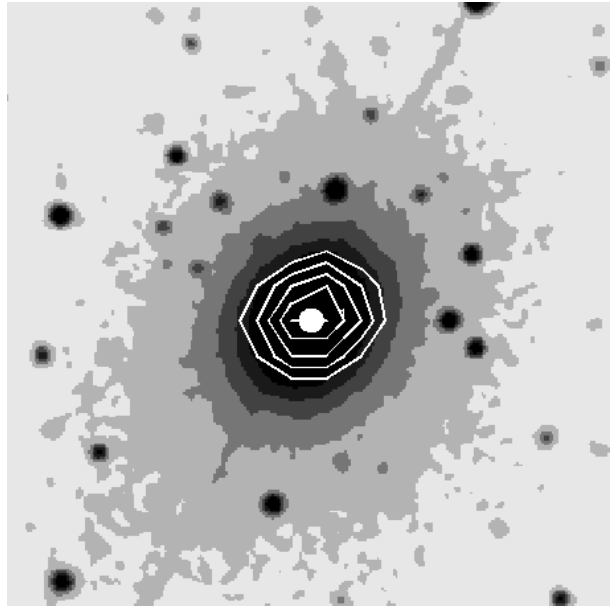} 
}}
\caption{The optical image and the radio iso-intensity contours of SHBL~J020014.8+312545=NGC~0777, a 
nearby elliptical galaxy and candidate HBL in the survey. The NVSS position
(filled symbol) of the flat spectrum radio source is consistent with that of nucleus of NGC~0777. 
}
\label{0777_fc}
\end{figure}

We have been looking for similar bright elliptical galaxies/low luminosity HBLs in the 
original sample of high \fxfr sources of Paper I (\fxfr $\ge 3\times 10^{-10}$ \ergj) 
among the objects that were excluded because their \aro was below the threshold value of 0.2.
We have found the five objects that are reported in Table 2 where columns 1,2 and 3 give the SHBL, the RASS 
and other names; column 4 gives the redshift, when available; column 5 is the visual apparent 
magnitude; column 6 is the X-ray luminosity in the Rosat band and column 7 gives the 
source classification from NED.
As an example Fig.~\ref{0777_fc} shows the optical image,
taken from the ESO on-line service, of one of the objects in Table 2 (1RXS~J020014.5+31254=NGC~0777)
with the radio iso-intensity contours and the precise position from the NVSS survey overlayed.
The flat/inverted radio spectrum ($\alpha_{\rm 1.4-2.38GHz}=-0.37\pm0.4$, $F(\nu)~\propto~\nu^{-\alpha}$) 
coincident with the nucleus of the galaxy and the \fxfr flux ratio make this source consistent 
with a high energy peaked BL Lac.

\subsection{High luminosity HBLs and the problem of redshift estimation}
\label{high_lum}

At the bright end of the BL Lac radio luminosity function, especially for the case of HBL objects where, 
for the same radio luminosity the optical output is much higher than in 
LBL objects (see Fig.~\ref{lbl_hbl}), the optical flux is totally dominated by the non-thermal, 
featureless nuclear emission which makes any redshift estimation a very difficult task.
Figure \ref{opt_sed} (panel d) is an example of a featureless optical spectrum belonging to a source
that could be a high luminosity HBL. Since there are 39 such objects in the sample
the fraction of high redshift (z \gsim 0.8) - high luminosity objects may be as high as $\sim 25\%$. 

In principle these objects could also be very bright nearby radio sources; 
however if that were the case their radio flux distribution would be heavily biased towards
high fluxes compared to that of the subsample of objects with a measured redshift. 
In Fig.~\ref{rflux_dist} the distributions of 1.4 GHz radio fluxes for BL Lacs with measured redshift 
(solid line) and for the subsample of objects where the redshift could not be measured due to the 
featureless nature of the optical spectrum (dotted line) are shown. As can be seen, the latter 
objects strongly peak at low radio flux values indicating that they cannot be nearby very high luminosity 
objects that out-shine the host galaxy but rather high redshift high luminosity sources. 

The only other alternative is that the luminosity of the host galaxies of these objects is 
extremely low contrary to the findings of \cite{Wur97} and \cite{Urry00}.
 
We conclude that it is highly probable that the featureless objects are high redshift-high 
luminosity High Energy Peaked BL Lacs. 

\subsection {Rejected broad-lined candidates}

The multi-frequency statistical selection criteria of the Sedentary survey have been estimated 
to be about 85\% efficient. This paragraph 
deals with those candidates that although satisfied all the selection conditions 
of Paper I were at some point excluded from the sample because they were either reported in the literature
as known emission line AGN or were not confirmed as BL Lacs by the optical identification process.  

Specifically, in 8 cases the candidate HBLs have been found to be known emission line AGN and were 
rejected in Paper I; other 3 candidates have been identified as 
emission line AGN by \cite{Bauer00} and \cite{Schwope00} after the publication of Paper I, while 
our optical spectroscopic campaign revealed 8 emission line AGN out of the 58 previously unclassified
candidates that were observed.

All together the classification of all candidates in the Sedentary Survey revealed that 
only 19 out of the original 163 objects in the ``HBL zone'' showed emission lines that 
are too strong for a source to be called BL Lacertae object according to the classification 
method of \cite{Marcha96}. 
The total level of contamination is therefore about $12\%$, well within the expected value 
of $\sim$15$\%$ (see Paper I).

The 19 rejected emission line AGNs are listed in Table 3 where column 1 gives the RASS name,
columns 2, 3 and 4 the fluxes in the X-ray (0.1-2.4 keV), radio (20 cm, from the 
NVSS survey) and the optical apparent $V$ magnitude (from APM and COSMOS, see Paper I) respectively, 
column 5 the redshift, column 6 and 7 give the radio luminosity and the \aox ; column 8 gives 
the reference for the optical identification. 

These X-ray sources are mostly nearby, low radio luminosity emission line AGN with \aro 
close to the selection threshold of 0.2. 
The radio emission of these sources is probably of non-nuclear origin and the X-ray emission may be 
unrelated to it.  

From previous surveys we know that FSRQs with low \fxfr are much more abundant than BL Lacs 
of similar \fxfr both at high (e.g. in the 1Jy sample \cite{Sti93d}, \cite{Sti91}) and lower 
radio flux (e.g. 50 mJy, in the DXRBS survey \cite{P97b}, \cite{L01}).  
Until recently X-ray strong (high \fxfr) FSRQ or HFSRQs (High energy peaked FSRQs) 
were instead thought to be very rare or altogether non existent; however \cite{P03} discovered 
that a relative large number of HFSRQs indeed exist, although none with \fxfr values as high 
as those required for inclusion in the sedentary survey. 

If the relative abundance of FSRQ and BL Lacs were independent of \fxfr (or, equivalently, on 
the position of the synchrotron peak energy) our survey, that includes 150 BL Lacs, should have 
detected several hundred extreme HFSRQs. Since we have possibly found only very few cases, all 
at relatively low redshifts and low radio power, we confirm with high statistical confidence 
that FSRQs very rarely (if at all) reach \fxfr values so high as to satisfy condition 2) in our survey 
definition criteria (\fxfr $\ge 3\times 10^{-10}$ \ergj).  In other words
the synchrotron component in broad lined Blazars do not reach peak energies as high as 
those found in HBL BL Lacs. 

\section{X-ray spectroscopy with \sax} 
 
The \sax X-ray Astronomy Satellite (see \cite{Boella97b} for a full description) 
was characterized by a very wide bandpass (0.1 $\sim 200 $ keV) and therefore 
particularly well suited to study the X-ray spectrum of bright extragalactic objects like HBL BL Lacs. 
\sax successfully operated for a period of six years, from May 1996 to May 2002, 
during which it accumulated a large data archive, unique in terms of bandwidth, that 
can be used to determine the broad band X-ray spectrum of these sources. 

A compilation of all the \sax spectral data for Blazars which were publicly available 
in March 2002 has been presented in \cite{Gio02c}. 
In the following we update this work concentrating on the subset of objects belonging to the Sedentary 
survey and we complete the dataset by adding all the observations carried out after March 2002.  

\begin{figure}[ht]
\centerline{ 
\vbox{ 
\includegraphics[width=6.5cm,angle=-90]{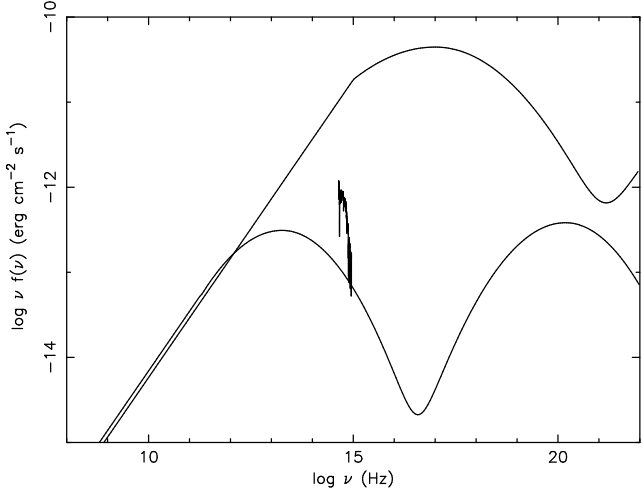} 
}} 
\caption{Typical HBL and LBL spectral energy distributions compared 
to  the SED of a typical giant elliptical galaxy. 
Note that the emission from the host galaxy is much more easily washed out 
by an HBL than by a LBL for the same radio flux.  
} 
\label{lbl_hbl}
\end{figure}

Twenty five objects of the Sedentary survey have been observed with \sax as part of different 
observing programs for a total of 50 pointings. \cite{Gio02c} considered 
several spectral models to fit the data and concluded that for HBL BL Lacs,
which very often exhibit continuous convex curvature in the X-ray band, the model that 
best matches the data is a logarithmic parabola of the type   
 
\begin{equation}
 F(E) = K~(E/E_*)^{-(a +b*Log(E/E_*))} ~~~~ (\mathrm{ph/cm^2/s/keV}) 
\end{equation}

\noindent 
This model is particularly appealing since it can describe broad-band spectral curvature with only three 
free parameters $K$, $a$ and $b$, just one more than a simple power law; 
in the following we assume $E_*= 1~keV$. 
\cite{Mas04} and \cite{Mas04b} carefully analyzed all the \sax observations of 
the bright HBL BL Lacs Mkn~421 and Mkn~501 and showed that the log-parabolic model is a good 
representation of the spectrum of these sources in all intensity states over the 
entire \sax energy bandpass. 
These authors also showed that the parameters of the logarithmic parabola model may provide 
information on the particle acceleration mechanism and can be used to easily calculate 
useful quantities such as an energy dependent photon index $\Gamma(E)$:
\begin{equation}
 \Gamma(E) = a~ +~ 2~ b~ Log(E/E_*) ~~~~~~~~.
\end{equation} 
The parameter $a$ is the photon index at the energy $E_*$, while $b$ measures the curvature of the 
parabola. The peak frequency $\nu_p=E_p/h$, corresponding to the maximum in the $\nu - \nu F(\nu)$ plot, 
is given by:
\begin{equation}
 E_p =  E_* ~ 10^{(2-a)/2b} ~~~ (\mathrm{keV})
\end{equation}
and the maximum value is:
\begin{equation}
\nu_p F(\nu_p) = (1.60\times10^{-9})~K~E_{*}^{2}~10^{(2-a)^2/4b} ~~{\rm erg/(cm^2~s)}~,
\end{equation}
(\cite{Mas04})).

The X-ray data analysis was carried out in a uniform way using the XANADU package (XIMAGE, 
XRONOS, XSPEC) and following the standard procedures described in the 
\sax documentation (http://www.asdc.asi.it/bepposax/).
Spectral fitting was carried out with XSPEC (v.~11.0) using the calibration files available from 
the \sax Science Data Center. 
We have used data from the LECS and MECS experiments and, whenever the source 
was bright enough we also included the high energy PDS data. 
The results are reported in Table 4 where column 1 gives the source name, 
column 2 gives the date of the \sax observation, column 3 gives the instruments used for the
analysis; columns 4, 5 and 6 
give the best fit parameters ($a$ $b$ and $K$) together with 1 sigma errors, column 7 gives 
the reduced $\chi^2$, column 8 gives 
the 2--10 keV flux in units of $10^{-11}$ \ergs and column 9 gives the peak energy derived using Eq.~(3).

\begin{figure}[ht]
\vbox{
\centerline{
\includegraphics[width=9cm] {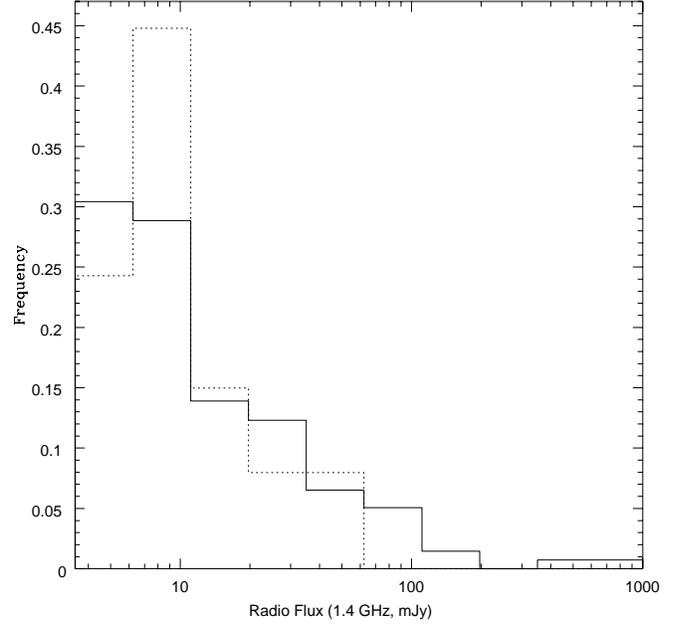} 
}}
\caption{The sky coverage corrected distributions of 1.4 GHz radio fluxes for BL Lacs with measured 
redshift (solid line) and for the subsample of objects where the redshift could not be measured due to the 
featureless nature of the optical spectrum (dotted line). 
}
\label{rflux_dist}
\end{figure}

\begin{figure}[ht]
\vbox{
\centerline{
\includegraphics[width=6.5cm,angle=-90] {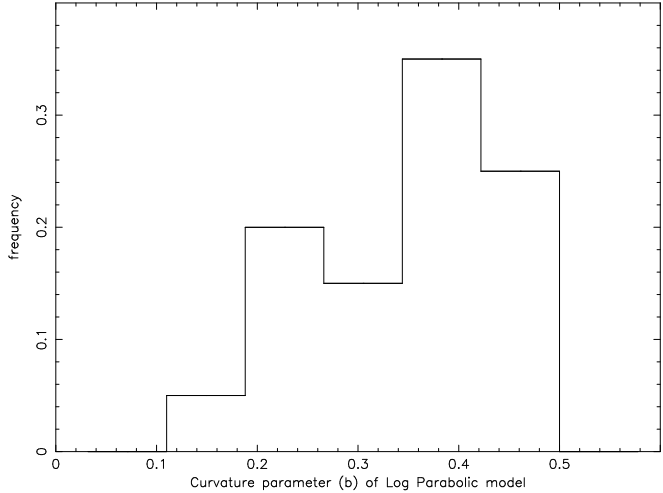} 
}}
\caption{The distributions of the curvature parameter ($b$) in the log parabolic model for
the subsample of objects (18) for which the parameter $b$ could be estimated with a statistical 
error less than 0.25}
\label{b_dist}
\end{figure}

As can be seen, in all but a few cases (see notes below), the wide band X-ray spectra of the objects 
in the Sedentary Survey  observed by \sax are satisfactorily described by the log-parabolic model. 
The curvature parameter ($b$) has been found to be positive, that is the spectrum is 
downward curved, in all sources with the exception of SHBL J020106.6+003401 and 
SHBLJ 123511.0-140322 (observation of July 1999) where the curvature estimation 
is very uncertain. The distribution of $b$ is shown in Fig.~\ref{b_dist} where it can be seen  
that the values of the curvature parameter measured with a reasonable accuracy cluster around values of 
$b \approx 0.4$ with a hint of a secondary peak at  $b \approx 0.2$. This second peak, however is
not statistically significant and larger samples would be needed to confirm or disprove it.

\begin{figure}[ht]
\vbox{
\centerline{
\includegraphics[width=6.5cm,angle=-90] {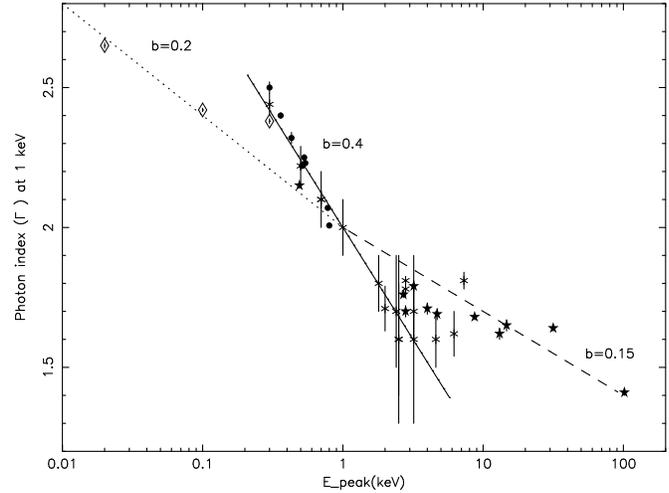} 
}}
\caption{The power law spectral index ($ \Gamma $) at 1 keV ($a$) plotted against the peak energy. 
A strong correlation is present for $E_p $ (as predicted by Eq.~(3) if $b$ is constant) 
values less than about 5 keV, above which  $\Gamma $ remains approximately constant. 
Crosses represent most of the objects 
listed in Table 3, diamonds are for PKS 2155-304, filled circles for Mkn~421 and filled stars 
for Mkn~501, which is not included in the sedentary survey because at the time of the RASS 
survey was in a very low state, but was added to this plot to extend the $E_p $ dynamical 
range to about 100 keV.
The solid, dotted and dashed lines represent the expected correlations for $b$=0.4, $b$=0.2 and 
$b$=0.15, respectively. 
}
\label{a_vs_epeak}
\end{figure}

The energy where the emitted power is maximum ($E_p$) can be calculated from $b$ and the measured 
photon index at 1 keV ($a$) using Eq.~(3). In Fig.~\ref{a_vs_epeak} the two quantities are plotted
against each other. 
In this plot  we have added the best fit values for Mkn~501 from \cite{Mas04b} to extend the $E_p$ 
range to about 100 keV. The plotted quantities should be on a straight line if the 
curvature parameter were the same in all objects. From Fig.~\ref{a_vs_epeak} we see that most 
of the sources actually lie along the $b=0.4$ line in most intermediate cases 
($0.1<E_{peak} < 5 $ keV) but significant deviations are apparent both at very low (PKS 2155-304) 
and at very high (Mkn~421 and Mkn~501 during flares) peak energies. 

\subsection {Notes on individual objects} 

\subsubsection{SHBL J020106.3+003401} 
 
This is the only object in the sample which shows concave (upward instead of downward) curvature in 
the \sax X-ray spectrum. Most of this curvature is due to a feature above 6 keV where the statistics 
are not good enough to allow us to perform a detailed analysis. Moreover this source is within 3 arc-minutes 
from the QSO SDSS~J020115.53+003135.1 which is clearly detected both in the \sax and in a Rosat PSPC 
image (although at a flux level lower than that of SHBL J020106.3+003401) that could contaminate 
the \sax data.

\subsubsection{SHBL J074405.6+743358} 
The logarithmic parabola fit to the X-ray spectrum of this source is not a good 
representation of the data.The reduced $\chi^2$ is 1.5 (17 d.o.f.) which is possibly 
due to the poor quality of the data and to a some flux excess above 4-5 keV.  

\onecolumn
\begin{table}[*t]
\caption{Results of \sax spectral fittings}
\begin{center}
\begin{tabular}{lcccccrrcc}
\hline
\\
Source name & \multicolumn{1}{c}{Observation}& Instrument(s)&
\multicolumn{4}{c}{Best fit parameters} &
\multicolumn{1}{c}{$\chi_r^2(\mathrm{d.o.f.}$) }&\multicolumn{1}{c}{flux$^{(a)}$} & $E_p$ \\
SHBL J & date& used$^{(*)}$ & $a$ & &$b$ &\multicolumn{1}{c}{$K$} &  && (keV)\\
\multicolumn{1}{c}{(1)}&(2) & (3) &\multicolumn{1}{c}{(4)} & & (5) & \multicolumn{1}{c}{(6)} & \multicolumn{1}{c}{(7)} & (8)& (9) \\

\hline
012308.7$+$342049 &03-01-99 & L M &$1.9\pm0.1$& &$0.2\pm0.1$&$(6.7\pm0.6)~10^{-3}$       &1.1(49)& 1.70 			& 1.8     \\
                  &02-02-99 & L M &$1.8\pm0.1$& &$0.4\pm0.1$&$(6.0\pm0.7)~10^{-3}$        &0.89(42)& 1.30 				& 1.8     \\
013632.5$+$390559 &09-01-01 & L M &$2.0\pm 0.1$& &$0.5\pm0.1$&$(6.6\pm0.4)~10^{-3}$        &1.11(44) &1.04  			& 1.0     \\
020106.3$+$003401 &16-08-96 & L M &$2.5\pm0.3$& &$-0.2\pm0.3$&$(1.6\pm0.7)~10^{-3}$        &1.25(20) &0.26  			& ...       \\
031951.9$+$184534 &15-01-97 & L M &$1.6\pm0.3$& &$0.4\pm0.2$&$(2.3\pm0.7)~10^{-3}$        & 0.75(40) &0.7  			& 3.2     \\
032613.9$+$022515 &20-01-98 & L M &$1.5\pm0.3$&&$0.7\pm0.4$ &$(1.2\pm0.6)~10^{-3}$        & 0.95(20) &0.3  			& 2.3     \\
034923.2$-$115927 &10-01-97 & L M &$1.7\pm0.2 $& &$0.4\pm0.2 $&$(2.4\pm0.6)~10^{-3}$        & 0.99(41) &0.6  		& 2.4     \\
041652.4$+$010524 &21-09-96 &  M  &$2.3\pm0.4$& &$0.3\pm0.3 $ &$(6.4\pm1.8)~10^{-3}$        & 0.76(33) &0.84 	& 0.3    \\
050939.0$-$040036 &11-02-99 & L M &$ 1.6\pm0.2 $& & $0.5\pm0.2 $ &$(2.0\pm0.5)~10^{-3}$        & 0.90(35) &  0.55     	& 2.5    \\
074405.6$+$743358 &29-10-96 & L M &$ 2.5\pm0.9 $& &  $0.2\pm0.7 $ &$(1.3\pm1.3)~10^{-3}$           & 1.50(17) &  0.14     	& 0.1    \\
093037.5$+$495025&25-11-98& L M & $1.7\pm0.1$&&$0.3\pm 0.1$&$(2.3\pm0.4)~10^{-3}$ &1.04(42)&0.66 				& 3.2    \\
103118.6$+$505335&01-05-97& L M & $2.1\pm0.1$&&$0.3\pm 0.1$ &$(6.2\pm0.3)~10^{-3}$ &0.75(42) &1.02 				& 0.7    \\
110337.6$-$232931         &04-01-97& L M P &$1.6\pm0.1$&&$0.3\pm 0.1$&$(11.7\pm0.8)~10^{-3}$ &1.22(82) &3.72 			& 4.6    \\
                          &19-06-98& L M   &$1.9\pm0.1$&&$0.3\pm 0.1$&$(11.7\pm0.6)~10^{-3}$&1.20(89) &2.51 			& 1.5    \\
110427.3$+$381231$^{(b)}$ &29-04-97& L M   &$2.25\pm 0.01$&&$0.45\pm 0.01$ & $(7.2\pm0.1)~10^{-2}$        &1.31(132)& 8.5 	& 0.5    \\
                          &30-04-97& L M   &$2.26\pm 0.01$ &&$ 0.47\pm 0.01$  &$(7.3\pm0.1)~10^{-2}$        &0.75(136) & 8.3 & 0.5    \\
                          &01-05-97& L M   & $2.23\pm 0.01$&&$0.43\pm 0.01$  &$(8.2\pm0.1)~10^{-2}$        &1.06(132) & 10.0 	& 0.5    \\
                          &02-05-97& L M   &$2.25\pm 0.01$&&$0.43\pm 0.02$  &$(10.2\pm0.2)~10^{-2}$        &0.86(127) & 12.3 	& 0.5    \\
                          &03-05-97& L M   &$2.32\pm 0.02$&&$0.44\pm 0.02$  &$(6.6\pm0.1)~10^{-2}$        &0.97(100) & 7.1 	& 0.4    \\
                          &04-05-97& L M   &$2.50\pm 0.02$&&$0.48\pm 0.02$  &$(5.0\pm0.1)~10^{-2}$        &1.16(100) & 4.1 	& 0.3    \\
                          &05-05-97& L M   &$2.40\pm 0.01$&&$0.45\pm 0.02$  &$(6.5\pm0.1)~10^{-2}$        &1.11(100) & 6.3 	& 0.4    \\
                          &21-04-98& L M P &$2.07\pm 0.004$&&$0.34\pm 0.006$  &$(18.8\pm0.1)~10^{-2}$   &1.19(171) & 31.0 & 0.8    \\
                          &23-04-98& L M P &$2.22\pm 0.004$&&$0.37\pm 0.007$  &$(13.5\pm0.1)~10^{-2}$  &1.02(169) & 17.8	& 0.5    \\
                          &22-06-98& L M P &$2.07\pm 0.007$&&$0.34\pm 0.008$  &$(14.4\pm0.1)~10^{-2}$        &1.26(150) & 23.9 & 0.8    \\
                          &04-05-99& L M P &$2.42\pm 0.004$&&$0.42\pm 0.005$  &$(11.4\pm0.1)~10^{-2}$        &1.11(141) & 11.0 & 0.3    \\
                          &26-04-00& L M P &$1.81\pm 0.002$&&$0.21\pm 0.002$  &$(22.7\pm0.1)~10^{-2}$        &1.06(152) & 62.5 & 2.8    \\
                          &09-05-00& L M P &$1.88\pm 0.003$&&$0.18\pm 0.003$  &$(19.4\pm0.1)~10^{-2}$        &1.34(149) & 49.5 & 2.2    \\
111706.3$+$201407 & 13-12-99 &  L M &$2.44\pm0.04$   & & $0.48\pm0.05$   & $(6.7\pm0.3)~10^{-3}$  & 1.43(54)  & 0.61 	& 0.3    \\
112048.0$+$421212 & 01-05-97 &  L M &$2.22\pm0.07$   & & $0.4\pm0.1$     & $(2.3\pm0.2)~10^{-3}$  & 0.40(25)  & 0.28 	& 0.5    \\
121158.6$+$224232 & 27-12-99 &  L M &$1.8\pm0.1$     & & $0.3\pm0.1$     & $(0.9\pm0.2)~10^{-3}$  & 1.10(43)  & 0.24 	& 2.2    \\
                  & 28-12-01 &  L M &$1.62\pm0.08$   & & $0.24\pm0.07$   & $(2.1\pm0.2)~10^{-3}$  & 1.03(54)  & 0.75 	& 6.2    \\
                  & 11-01-02 &  L M &$1.4\pm0.2$     & & $0.7\pm0.2$     & $(1.7\pm0.3)~10^{-3}$  & 1.45(41)  & 0.52 	& 2.7    \\
122121.9$+$301037 & 12-07-99 &  L M &$2.10\pm0.03$   & & $0.37\pm0.03$   & $(9.5\pm0.3)~10^{-3}$  & 0.78(88)  & 1.48 	& 0.7    \\
123511.0$-$140322 & 27-06-99 &  L M &$1.4\pm0.3$     & & $0.9\pm0.4$     & $(0.8\pm0.2)~10^{-3}$  & 0.56(19)  & 0.19 	& 2.2    \\
                  & 16-07-99 &  L M &$2.6\pm0.2$     & & $-0.4\pm0.3$    & $(1.0\pm0.2)~10^{-3}$  & 0.46(19)  & 0.15 	& ...      \\
125731.9$+$241240 & 20-06-98 &  L M &$1.6\pm0.3$     & & $0.5\pm0.3$     & $(3.8\pm1.2)~10^{-3}$  & 0.84(38)  & 1.11 	& 2.5    \\
141756.1$+$254356 & 13-07-00 &  L M &$1.71\pm0.08$   & & $0.47\pm0.07$   & $(5.2\pm0.4)~10^{-3}$  & 1.15(49)  & 1.26 	& 2.0    \\
                  & 23-07-00 &  L M &$1.5\pm0.2$     & & $0.7\pm0.2$     & $(3.4\pm0.9)~10^{-3}$  & 0.84(36)  & 0.88 	& 2.3    \\
                  & 27-07-00 &  L M &$1.85\pm0.08$   & & $0.30\pm0.07$   & $(4.0\pm0.4)~10^{-3}$  & 1.25(49)  & 0.93 	& 1.8    \\
142832.6$+$424024 & 08-02-99 &  L M &$1.81\pm0.03$   & & $0.11\pm0.03$   & $(6.6\pm0.3)~10^{-3}$  & 0.98(87)  & 2.03 	& 7.3    \\
151747.4$+$652523 & 05-03-97 &  L M &$2.0\pm0.2$     & & $0.4\pm0.2$     & $(5.5\pm0.8)~10^{-3}$  & 1.30(39)  & 0.98 	& 1.0    \\
153500.9$+$532037 & 13-02-99 &  L M &$2.0\pm0.3$     & & $0.4\pm0.3$     & $(1.4\pm0.5)~10^{-3}$  & 0.89(38)  & 0.25 	& 1.0    \\
215852.0$-$301331 & 20-11-96 &L M P &$2.42\pm0.006$ & & $0.23\pm0.01$ & $(49.3\pm0.4)~10^{-3}$ & 1.22(130) & 5.59 		& 0.1    \\
                  & 22-11-97 &L M P &$2.38\pm0.008$ & & $0.36\pm0.01$   & $(77.6\pm0.8)~10^{-3}$ & 1.28(108) & 8.28 	& 0.3    \\
                  & 04-11-99 &L M   &$2.65\pm0.009$ & & $0.20\pm0.01$   & $(28.9\pm0.4)~10^{-3}$ & 1.16(103) & 2.47 	& 0.02    \\
235907.9$-$303739 & 21-06-98 &L M P &$1.78\pm0.03$   & & $0.25\pm0.03$   & $(8.8\pm0.3)~10^{-3}$  & 1.35(91)  & 2.43 	& 2.8    \\
\hline
\end{tabular}
\end{center}
 $^{(a)}$ X-ray flux in the 2--10 keV band in units of $10^{-11}$ \ergs \\
 $^{(b)}$ Spectral parameters taken from \cite{Mas04}  \\
 $^{(*)}$ L=LECS; M=MECS; P=PDS
\vspace*{6truecm}
\end{table}
\twocolumn

\section{Broad-Band Spectral Energy Distributions and Synchrotron Peaks}

One of the main motivations of the Sedentary Survey was the selection of 
a sizable {\it radio flux limited} sample of HBL BL Lacs, a type of sources 
that in the past have been discovered almost exclusively in X-ray surveys. 

The very high X-ray to radio flux ratio that characterizes these objects is thought to 
be the result of synchrotron radiation extending to X-ray or even higher energies.  
To verify that this is indeed the case we have constructed the broad-band SED 
(in the usual $Log(\nu F(\nu))~vs~Log(\nu) $ space) combining the \sax X-ray data 
(de-reddened using the cross sections of 
\cite{Mor83} setting the amount of absorbing material ($N_H$) equal to the 
Galactic value along the line of sight), with the (non-simultaneous) 
optical spectroscopy data from the sedentary identification campaign when available, and 
with non-simultaneous multi-frequency literature data taken from NED, from 
the NVSS 20 cm survey (\cite{Con98}), 
the Two Micron All Sky Survey (2MASS, \cite{Skru95}), the Sloan Digital Sky Survey
(SDSS, \cite{Sloan01}), the GSC2 catalog \cite{lasker95, McLean00} and the 
Rosat All Sky Survey (\cite{Vog99}).

\subsection{2MASS and GSC2 fluxes} 
 
The Two Micron All Sky Survey (2MASS) covers the full sky 
at near infra-red frequencies and allows us to add three flux measurements to 
our SEDs, at least for the brightest sources in our Survey. Indeed 
for all objects brighter that  $V \sim 17.5$ a counterpart in the 2MASS Point 
Source Catalog (\cite{cutri03}) has been found.
 
We have converted $J$ $H$ and $Ks$ magnitudes from the 2MASS survey into 
monochromatic fluxes at  1.24, 1.66 and 2.16 microns assuming the following 
calibration (e.g. \cite{Coh03})  \\

$f(1.24\mu)=10^{-0.4(m_J-15.51)}mJy$ 
 
$f(1.66\mu)=10^{-0.4(m_H-15.03)}mJy$ 
 
$f(2.16\mu)=10^{-0.4(m_{Ks}-14.56)}mJy$ \\

\noindent GSC2 $F$ and $J$ magnitudes have been converted as follows: \\

$f(4.55 10^{14} Hz)=10^{-0.4(m_F-16.16)}mJy$ 

$f(6.00 10^{14} Hz)=10^{-0.4(m_J-16.43)}mJy$ \\

\noindent all magnitudes have been de-reddened according to the prescriptions
of \cite{Car89}.

\subsection{SED shapes} 

\begin{figure}[ht]
\begin{center}
\vbox{
  \includegraphics[width=6.0cm,angle=-90] {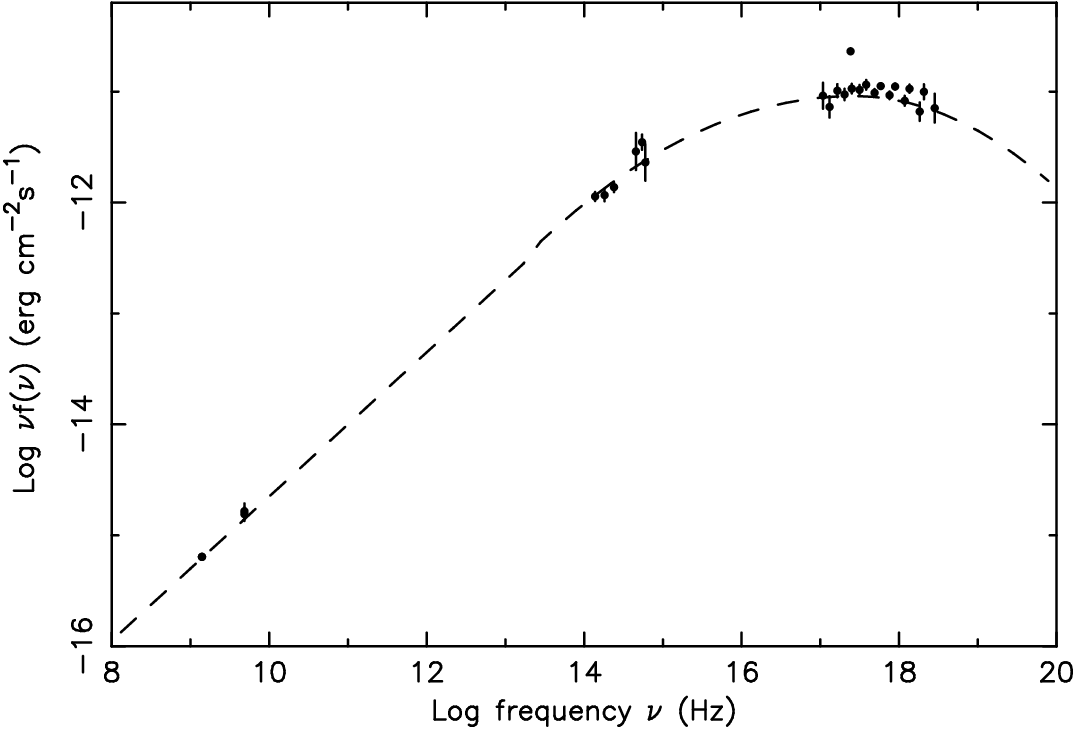}
  \caption{Spectral Energy Distribution of SHBL~J012308.7+342049}
  \label{sed_first}

  \includegraphics[width=6.0cm,angle=-90] {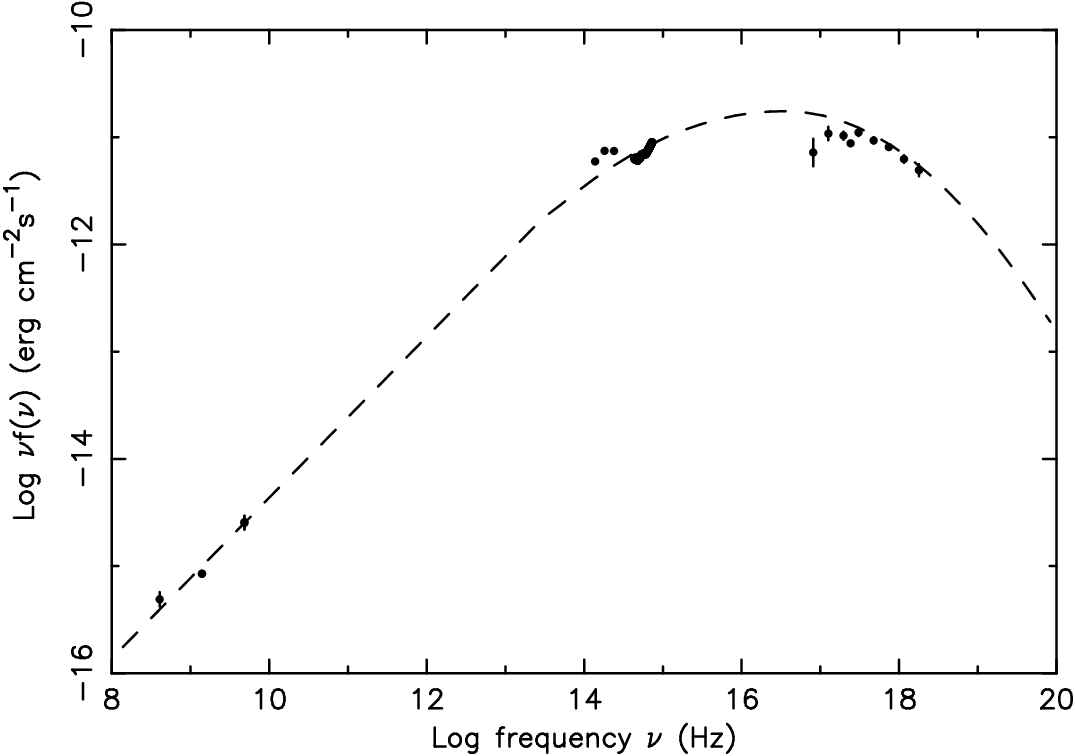}
  \caption{Spectral Energy Distribution of SHBL~J013632.5+390559}

  \includegraphics[width=6.0cm,angle=-90] {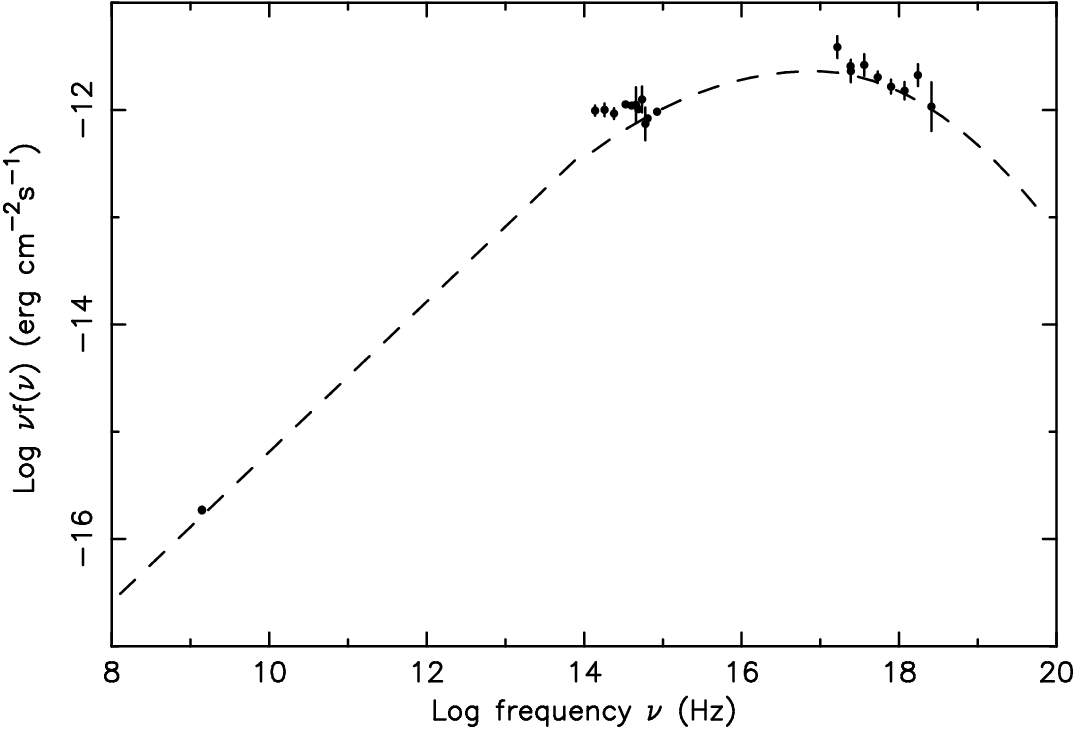}
  \caption{Spectral Energy Distribution of SHBL~J020106.3+003401}

} 
\end{center}
\end{figure}

\begin{figure}[ht]
\begin{center}
\vbox{

\includegraphics[width=6.0cm,angle=-90]{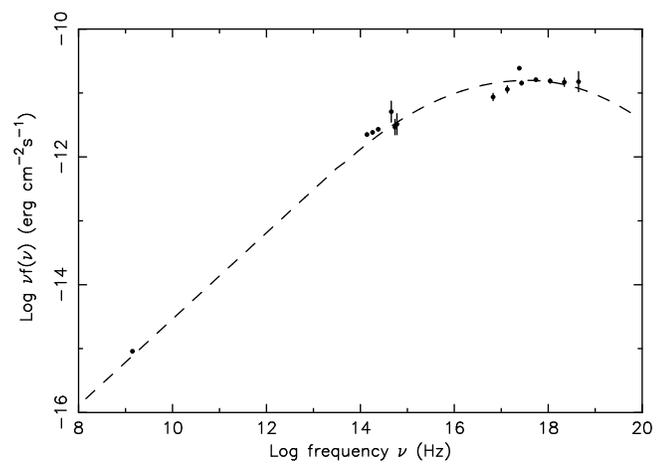}
\caption{Spectral Energy Distribution of SHBLJ~235907.9$-$303739}
\label{sed_last}

}
\end{center}
\end{figure}

The SEDs of all objects for which \sax data are available are plotted in 
Figs.~\ref{sed_first} through \ref{sed_last}; for reasons of brevity the SED 
for the remaining objects, which include only a few points, are not plotted here 
but are available on-line at the following web site  

\begin{center}
http://www.asdc.asi.it/sedentary/ 
\end{center}

Although the data are not simultaneous, 
the $\nu f(\nu)$ dynamical range ($\approx$ a factor of 1000) is much larger than the 
expected variability and allows us to characterize the broad band spectrum of HBL BL Lacs
comparing the observed energy distribution to the expectations of a homogeneous Synchrotron 
Self-Compton (SSC) model adapted from Tavecchio, Maraschi \& Ghisellini (1998). 
This model assumes that radiation is produced by a population of relativistic electrons 
emitting synchrotron radiation in a single zone of a jet that is moving at relativistic 
speed and at a small angle with respect to the line of sight. These photons are subsequently scattered 
by the same electrons to higher energies via the inverse Compton process (ignoring 
the Comptonization of external photons, which only affects the SED at  $\gamma$-ray energies; 
e.g., \cite{Ghi98}). The physical parameters that define the model are the jet 
radius, the Doppler factor, the magnetic field B, and four spectral parameters of the electron 
population, assumed to follow a power-law distribution which turns into a log parabolic 
distribution above a given 
energy: the normalization, the two spectral slopes, and the break energy. The Klein-Nishina 
cross section is used in the computation of the Compton scattering.

In a number of cases (e.g. SHBLJ012308.7+342049, SHBLJ041652.4+010524 SHBLJ103118.6+505335 etc.) 
the X-ray spectral data is consistent with being the smooth extrapolation of the same synchrotron 
emission observed at radio and optical frequencies, as predicted by a homogeneous 
single zone SSC model. In other cases (e.g. SHBLJ0301951.9+184534, SHBLJ032613.9+022515, 
SHBLJ142832.6+424024 etc.), although the X-ray intensity is approximately located on the 
prediction of homogeneous SSC models, the low 
energy back-extrapolation of the \sax data clearly falls below the observed optical emission. 
This has been been also noted in the detailed analysis of several \sax observations of 
Mkn~421 and Mkn~501 (\cite{Mas04}, \cite{Mas04b}) who interpreted this as due to the
superposition of different emission components, the most energetic one possibly due to 
highly energetic electrons located in a very compact region inside the jet. Because of its 
small size the power output of this high energy component is only a fraction of the overall 
emission at optical frequencies but it becomes the dominant emission at very high energies where 
all other components have dropped well below their peak emission.  

\section{Discussion}

We have presented the final, cleaned sample of the ``Sedentary Survey of extreme
HBL BL Lacs'' which comprises 150 sources and is currently the deepest, largest, 
statistically complete and 100\% identified flux-limited sample of BL Lacertae objects.

By means of multi-frequency literature data, our own optical spectroscopic observations 
and of wide band X-ray data from the \sax public archive we have investigated the local 
and the broad-band spectral properties of the sample.

We have found that:

\begin{enumerate} 

\item The completion of the optical spectroscopy campaign, which led to the identification 
of all the objects in the survey (Paper III), fully confirmed the original assumption 
that a very high percentage of the sources in the initial sample were indeed BL Lac objects. 
In fact, 50 out of 58 previously unidentified candidates observed by us, or 86\%, turned 
out to be BL Lacs, a similar percentage was found in the sources identified as part of
different programs.  
This result validates the correctness of the selection method and the robustness of the 
preliminary results reported in \cite{Gio99} and  \cite{Perri02}. 
The revised results on the statistical properties of the sample, including LogN-LogS, luminosity 
function and cosmological evolution, will be presented in a separate paper (Paper IV).

\item A significant fraction of the sources ($\sim 25 \%$) displays a smooth non-thermal
optical continuum without features due to either narrow emission lines or to the host galaxy.
No redshift or luminosity can therefore be derived for these objects. However, since 
this condition only occurs when the nuclear non-thermal emission is much brighter than 
that of the host galaxy, a lower limit to the luminosity can be obtained setting the 
luminosity of the BL Lac equal to that of the host galaxy which is typically a giant 
elliptical with approximately constant absolute magnitude (e.g. \cite{Wur97}, \cite{Urry00}). 
Unless these sources are all associated with (never observed) very low luminosity hosts 
they must be the most luminous objects in the survey. An independent indication that 
these sources are likely to be high redshift objects is provided by the progression of the 
dominance of the non-thermal over the galaxian component with redshift shown in Fig.~\ref{opt_sed} 
and by the distribution of radio fluxes shown 
in Fig.~\ref{rflux_dist}.

We conclude that the 25\% of featureless sources in the sample are very likely 
intrinsically bright and therefore probably represent plentiful examples of the 
yet unreported {\it high radio luminosity--high energy peaked} BL Lacs. However, 
how luminous it remains to be determined. The existence of these sources would be at variance 
with the claimed inverse proportionality between radio power and synchrotron peak energy known 
as the ``Blazar sequence''. 

This luminosity sequence (\cite{Fos98}), which was interpreted as the direct consequence of 
differential cooling efficiencies in low and high radio power objects (\cite{Ghi98}), was based 
on the absence of {\it low radio power - low \nupeak} and of {\it high radio power - 
high \nupeak} objects in a composite sample of BL Lacs and FSRQs detected in X-ray and radio 
flux limited surveys which, although each complete above its flux limit, probed widely different 
radio luminosity regimes.

The ``Blazar sequence'' has been recently tested using new deeper, larger and more homogeneous 
(i.e. with a single flux limit) samples of Blazars. The existence of low-radio luminosity-low 
\nupeak objects has been demonstrated by \cite{P03} and \cite{Cac04} who also showed that these
are core-dominated radio sources just like the other BL Lacs and that their low radio
power cannot be explained as the consequence of large orientation angles. In addition, the
existence of high radio luminosity-high \nupeak sources has been discussed
by \cite{Gio02b} on the basis of preliminary results of the Sedentary Survey.

\item The relative abundance of FSRQ and BL Lacs of any \fxfr flux ratio is approximately 
of three-four FSRQ for each BL Lac (e.g \cite{Sti91}, \cite{L01}). 
Although the selection criteria of the Sedentary Survey do not preclude the detection of 
FSRQs, we have found 150 extreme HBL BL Lacs and almost no FSRQs, instead of several hundred
if the relative abundance of strongly lined and line-less Blazars were independent of \fxfr. 
Although a number of intermediate FSRQs have recently been found by \cite{Pad02b}, \cite{P03} it is clear 
that broad lined Blazars do not reach the very high synchrotron peak energies of HBL BL Lacs.

\item A few bright nearby elliptical galaxies have been found below the radio loud-radio quiet 
border (\aox = 0.2, see Table 2). 
These objects would normally be labelled ``radio galaxies'' or ``normal ellipticals'' and 
therefore removed from BL Lacs samples. However, these objects, which generally appear as 
point-like sources in the NVSS, may also be faint HBL BL Lacs whose 
non-thermal optical component is simply below the emission from the host galaxy. This situation must 
be expected towards the low end of the radio luminosity function.

In addition, strong flux variability, one of the defining properties of BL Lacs, and 
the position of the synchrotron peak (see Fig.~\ref{lbl_hbl}), both have a strong influence on 
the ratio between the galactic and non-thermal optical flux, blurring the border between 
BL Lacs and radio galaxies. For example a borderline source that varied its non-thermal flux 
by a significant amount would be classified as a BL Lac when in a high state and as a 
radio galaxy when in the low state. Also, as shown in Fig.~\ref{lbl_hbl} two BL Lacs with 
identical radio flux but with synchrotron peaks at low and high energy (e.g. LBL and
HBL) would also be classified differently depending on the position of the synchrotron peak. 


\item 
\cite{Mas04} and \cite{Mas04b} have shown that the wide band X-ray spectrum of the bright 
HBL BL Lacs like Mkn~421 and Mkn~501 can be satisfactorily described by a log-parabolic model 
in all intensity states (see Eq.~(1)). We have then fit this model 
to the X-ray spectrum of all the objects in the survey that have been observed by \sax
obtaining statistically acceptable fits in all but a few cases (see Table 4).
The curvature parameter ($b$) has been found to be positive (i.e. the spectrum is 
downward curved) in almost all sources.

\item The broad band SEDs confirm that these objects are HBLs, that is their  
synchrotron emission extends to very high energies, sometimes well into the X-ray 
band. The X-ray data are generally located on the extrapolation for the radio 
and optical spectral distribution as predicted by homogeneous SSC models with the synchrotron power 
peaking at X-ray frequencies. However, the local strong X-ray curvature in a number of objects 
is not consistent with a simple back-extrapolation to optical frequencies suggesting that there 
may be more than one emission component. 

\item The particularly energetic physical conditions that are necessary to produce 
the very high energy synchrotron photons observed imply that the corresponding Inverse Compton 
radiation must reach energies close to or within the TeV band. Indeed, even 
the source with the lowest synchrotron peak energy reported in Table 3 (i.e. PKS2155-304, 
\nupeak = 0.02-0.3 keV) has been detected at TeV frequencies (\cite{Chad99},\cite{Hin03}). 
It is therefore natural to expect that many of the sources in the Sedentary survey are TeV 
emitters and that the brightest and closest ones may be detectable by the 
present generation of Cherenkov telescopes, especially during flares. 
However, despite this obvious prediction only 3 (Mkn421=SHBL J110427.3+381231, 
PKS2155-304=SHBL J215852.0-301331 and H1426+428=SHBL J142832.6+424024) of the 6 presently 
established TeV BL Lacs are actually included in our survey. Mkn~501, 1ES2344+514 and 1ES1959+650, 
the remaining TeV Blazars, all displayed synchrotron peak energies well into the hard X-ray 
band (\cite{Mas04b}, \cite{Gio00}, \cite{Kraw04}) during strong outbursts but never reached (not even during 
the strongest flare) a soft X-ray flux ratio high enough to meet the \fxfr condition necessary 
to be part of the Sedentary survey. 
Most of the X-ray variations in these sources were in fact confined to the hard X-ray band, close 
to the maximum of their synchrotron power and above the Rosat bandpass. This implies that synchrotron 
peak energies in the hard X-ray band are not necessarily located on the smooth extrapolation 
of the lower energies spectrum, but may be due to additional, very energetic emission components that 
only emerge above the Rosat X-ray band. The existence of such sub-components is consistent with the 
observed difference between the shape of the overall SED and the local X-ray spectral curvature 
in many objects (see Figs.~\ref{sed_0301p18}, \ref{sed_0326p02}, \ref{sed_0509m04}, \ref{sed_mkn421},
\ref{sed_1257p24}, \ref{sed_1428p42} and \ref{sed_1517p65}). 

It is likely that TeV emission is associated with these components, some of which may not 
always be detectable at soft X-ray frequencies because they could be outshone by the main or 
by some other less energetic synchrotron component. This would be consistent with the existence
of the recently reported ``X-ray orphan'' TeV flares (\cite{Kraw04}), that is TeV flare emission 
not correlated to soft X-ray flares and that this lack of correlation should be expected in other 
objects. Simultaneous sensitive hard-X-ray observations are obviously desirable and could be 
achieved in the short term by organizing TeV observations with the Swift spacecraft and on the 
medium term with the next generation of hard X-ray imaging telescopes.

\end{enumerate}

\begin{acknowledgements} 

This work is partly based on 

\noindent \sax X-ray data taken from the \sax public archive hosted at 
the ASI Science Data Center (ASDC), Frascati, Italy, and on \\ 

\noindent Optical spectroscopy observations performed at the
European Southern Observatory, La Silla, Chile, (Proposals ESO n. 67.B-0222(A), 71.B-0582(A)
and 71.B-0582(B)), Telescopio Nazionale Galileo, La Palma, Canarian Islands
(proposals AOT5/02A, AOT6/02B, AOT7/03A) and Kitt Peak National Observatory.
 
\noindent This research has also made use of data taken from the following on-line services 

the NASA/IPAC Extragalactic Database (NED) 

the ESO on-line Digitized Sky Survey 

the Two Micron All Sky Survey, a joint project of the University of Massachusetts and IPAC,
funded by NASA and NSF  

the Sloan Digital Sky Survey archive, which is funded by the Alfred P. Sloan Foundation

the Guide Star Catalog-II, which is a joint project of the Space Telescope Science Institute and
the Osservatorio Astronomico di Torino. \\
The authors are grateful to E. Massaro for useful suggestions.

\end{acknowledgements}

\newpage
\begin{figure}[ht]
\begin{center}
\vbox{

  \includegraphics[width=6.0cm,angle=-90] {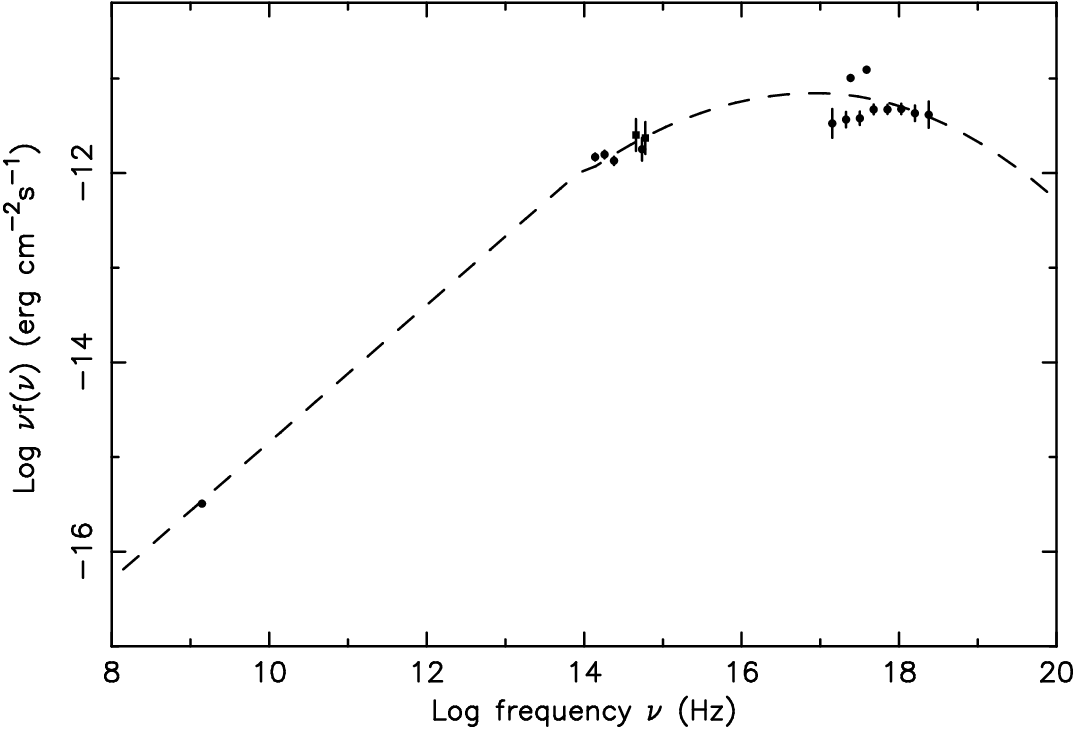}
  \caption{Spectral Energy Distribution of SHBL~J031951.9+184534}
  \label{sed_0301p18}

  \includegraphics[width=6.0cm,angle=-90] {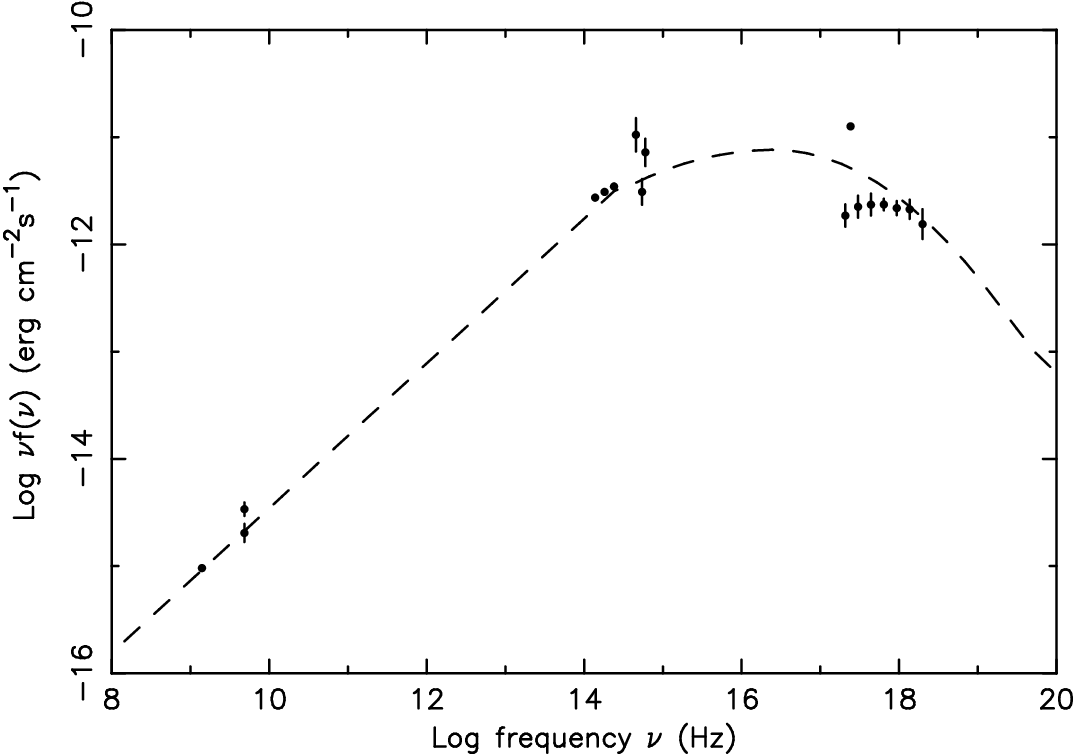}
  \caption{Spectral Energy Distribution of SHBL~J032613.9+022515}
  \label{sed_0326p02}

  \includegraphics[width=6.0cm,angle=-90]{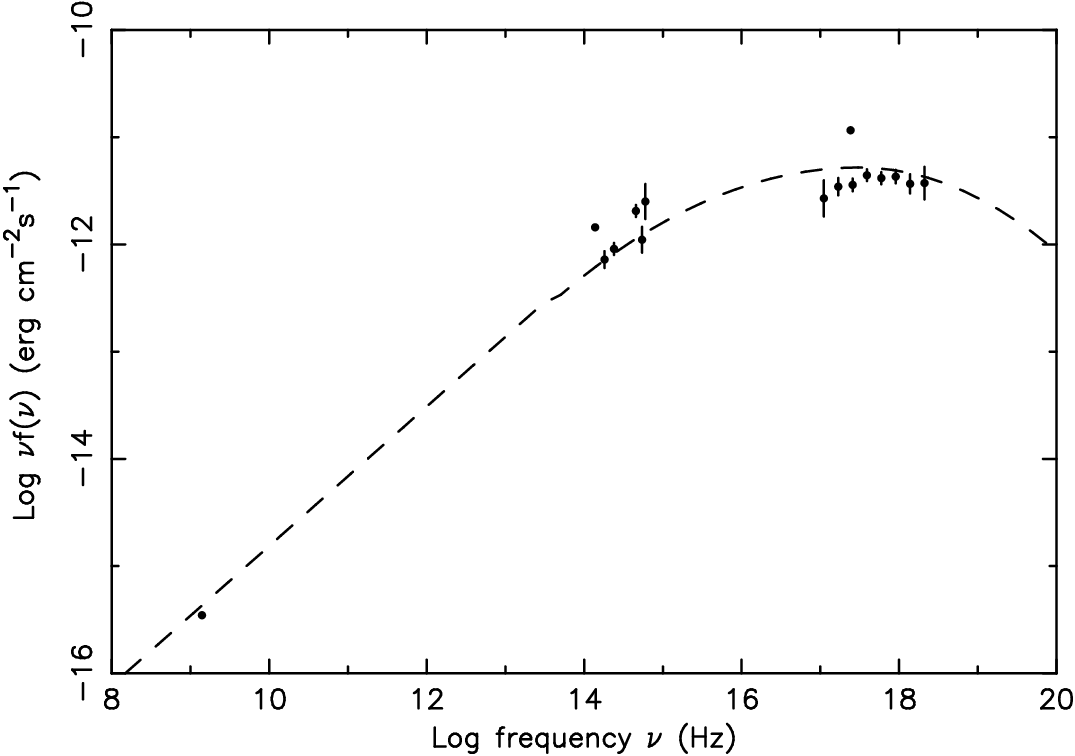}
  \caption{Spectral Energy Distribution of SHBL~J034923.2$-$115927}

}
\end{center}
\end{figure}

\begin{figure}[ht]
\begin{center}
\vbox{

  \includegraphics[width=6.0cm,angle=-90] {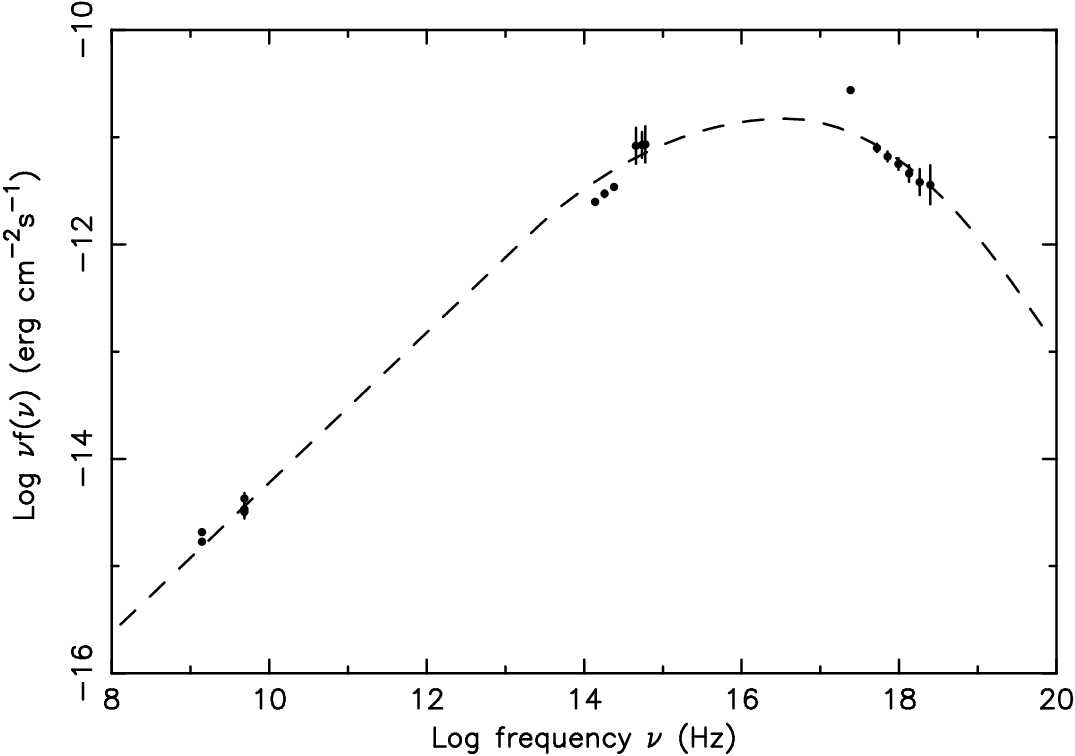}
  \caption{Spectral Energy Distribution of SHBL~J041652.4+010524}

  \includegraphics[width=6.0cm,angle=-90] {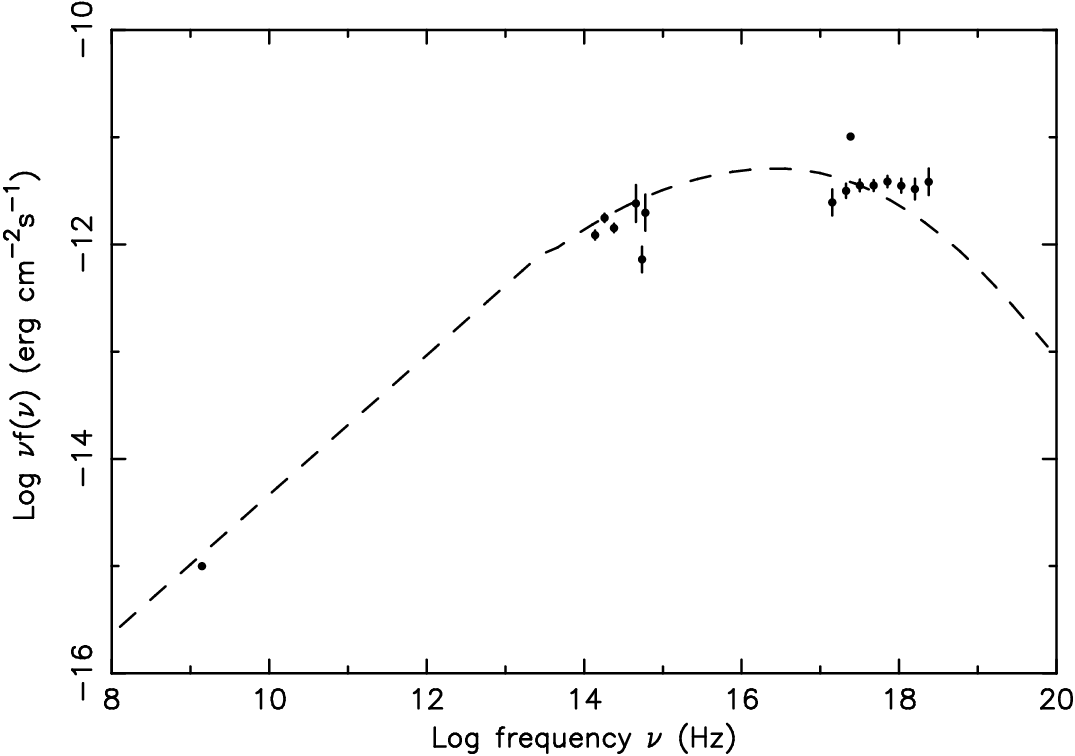}
  \caption{Spectral Energy Distribution of SHBL~J050939.0$-$040036}
  \label{sed_0509m04}

  \includegraphics[width=6.0cm,angle=-90]{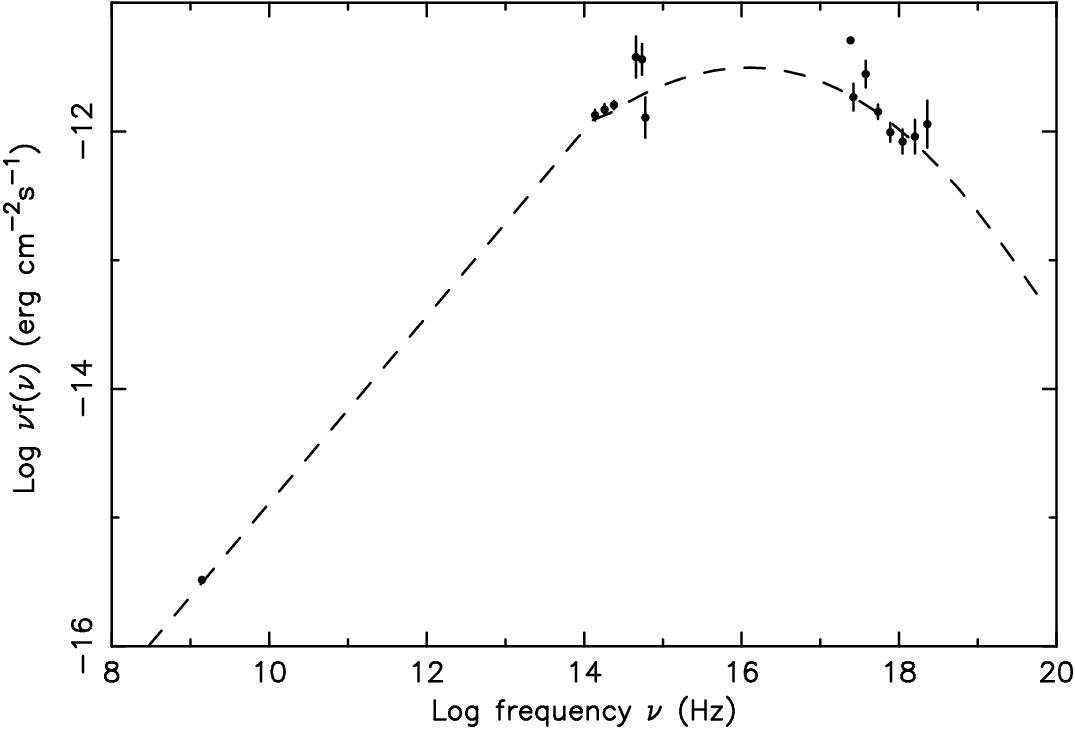}
  \caption{Spectral Energy Distribution of SHBL~J074405.6+743358}}

\end{center}
\end{figure}

\begin{figure}[ht]
\begin{center}
\vbox{

\includegraphics[width=6.0cm,angle=-90]{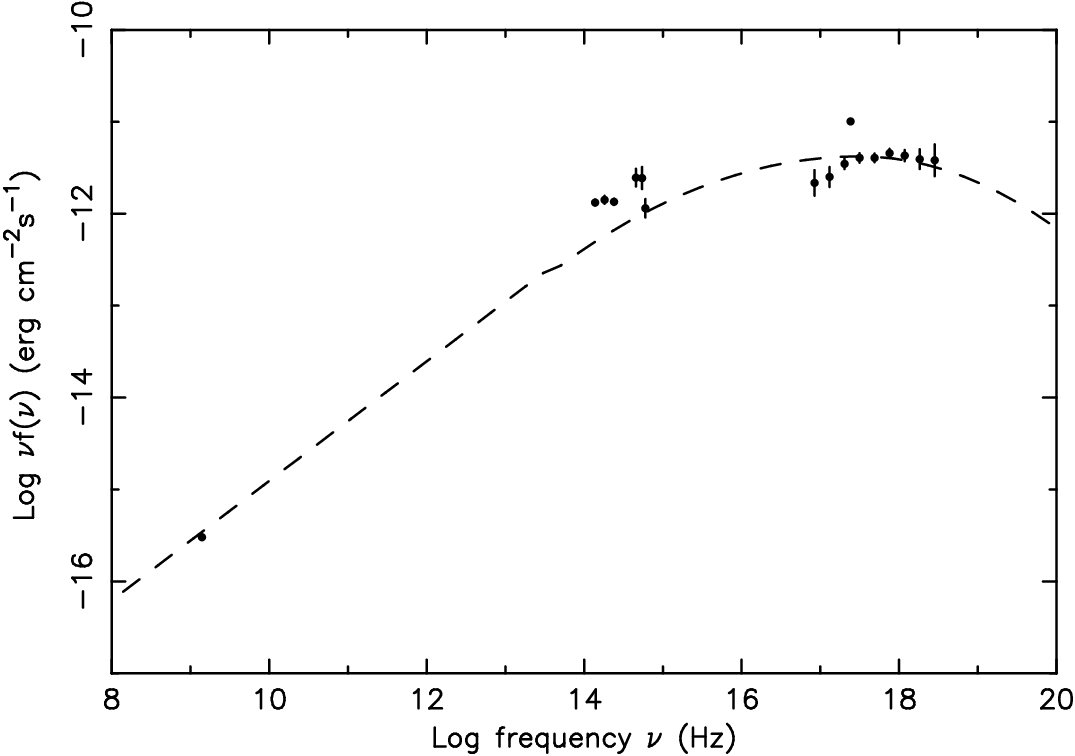}
\caption{Spectral Energy Distribution of SHBL~J093037.5+495025}

\includegraphics[width=6.0cm,angle=-90]{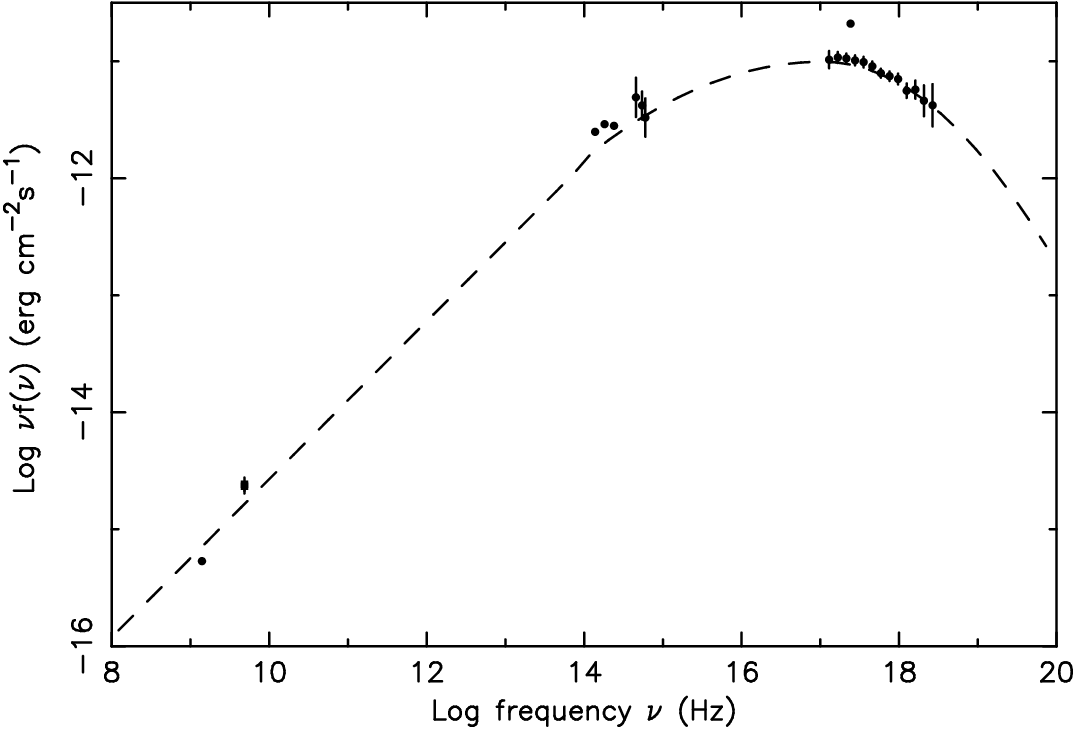}
\caption{Spectral Energy Distribution of SHBL~J103118.6+505335}

\includegraphics[width=6.0cm,angle=-90]{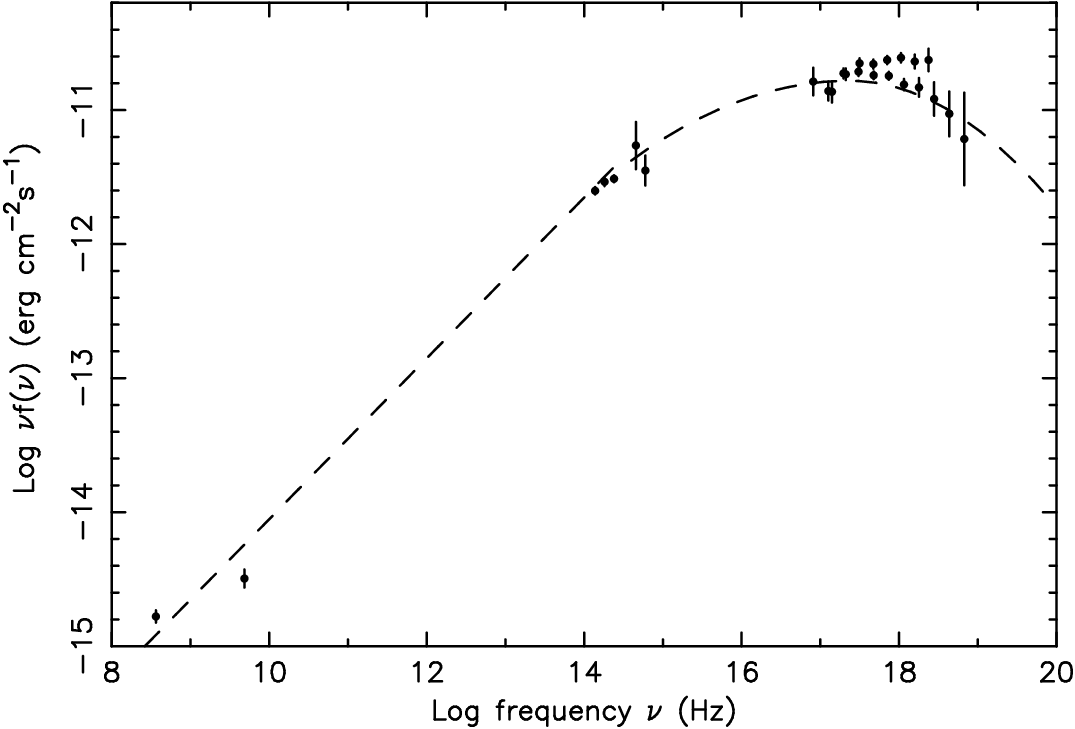}
\caption{Spectral Energy Distribution of SHBL~J110337.6$-$232931}

}
\end{center}
\end{figure}

\begin{figure}[ht]
\begin{center}
\vbox{

\includegraphics[width=6.0cm,angle=-90]{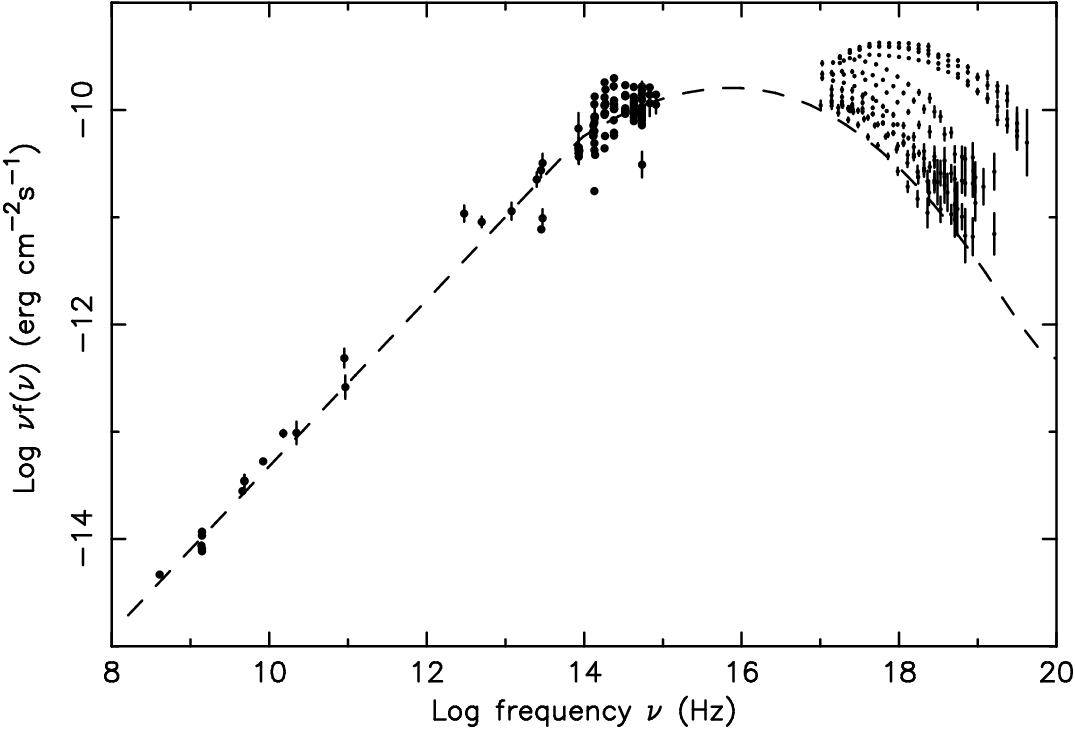}
\caption{Spectral Energy Distribution of SHBLJ~110427.3+381231=MKN 421}
\label{sed_mkn421}

\includegraphics[width=6.0cm,angle=-90]{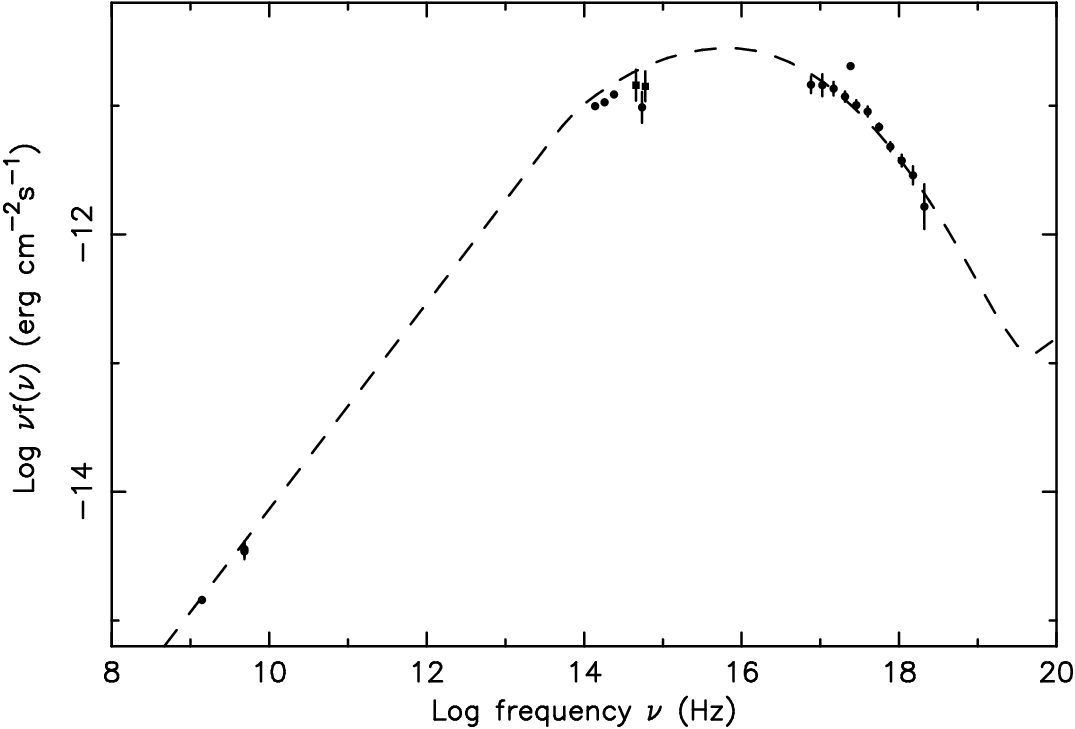}
\caption{Spectral Energy Distribution of SHBLJ~111706.3+201407}

\includegraphics[width=6.0cm,angle=-90]{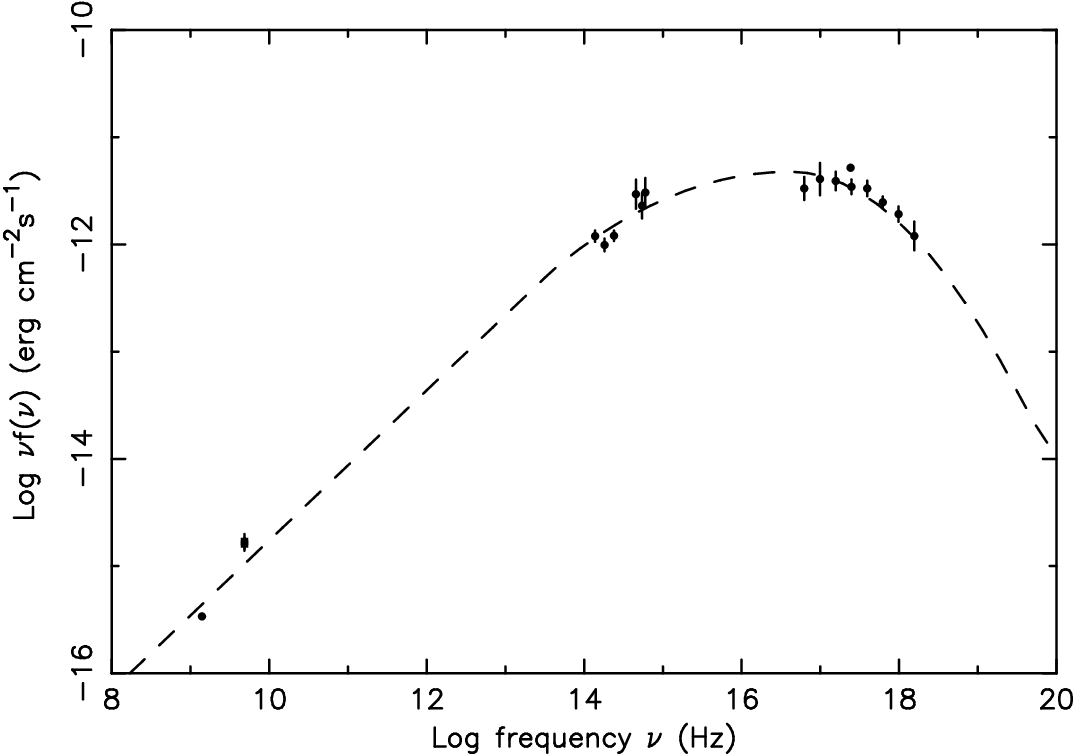}
\caption{Spectral Energy Distribution of SHBLJ~112048.0+421212}

}
\end{center}
\end{figure}

\begin{figure}[ht]
\begin{center}
\vbox{

\includegraphics[width=6.0cm,angle=-90]{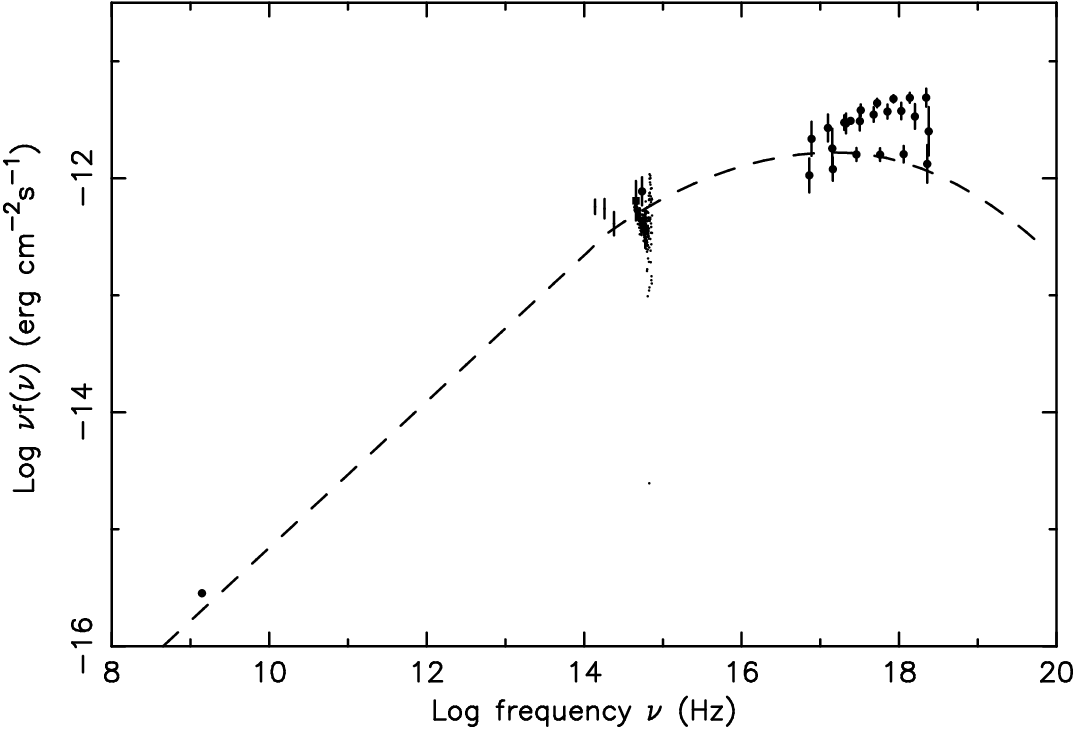}
\caption{Spectral Energy Distribution of SHBLJ~121158.6+224232}

\includegraphics[width=6.0cm,angle=-90]{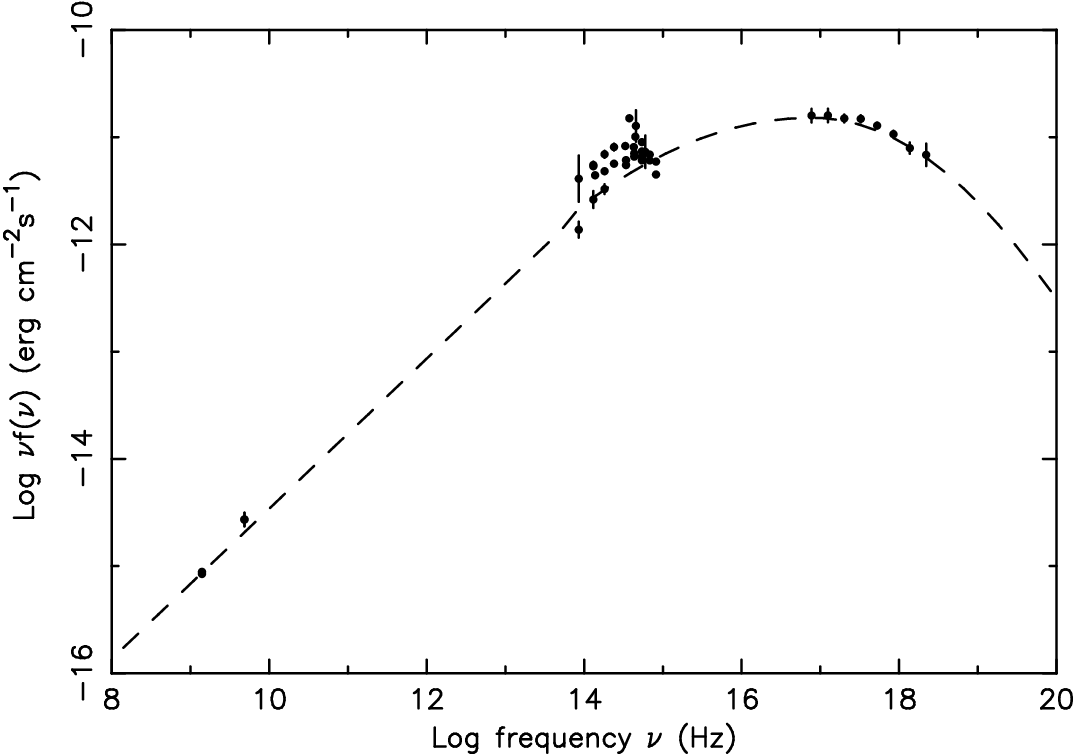}
\caption{Spectral Energy Distribution of SHBLJ~122121.9+301037}

\includegraphics[width=6.0cm,angle=-90]{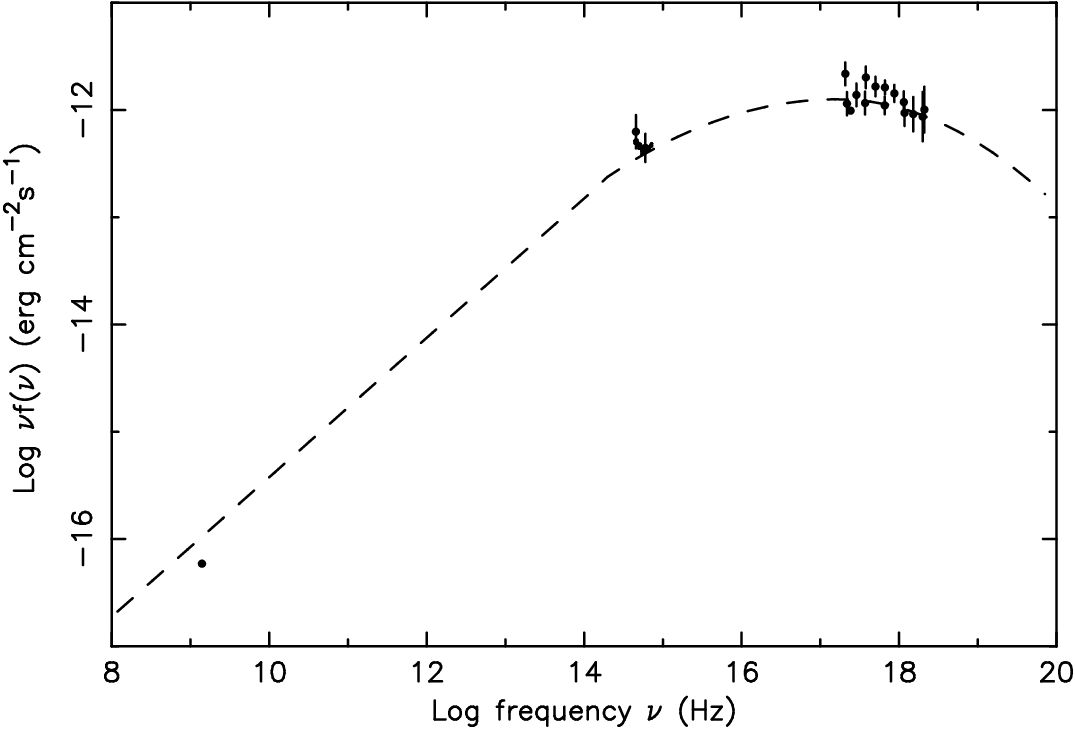}
\caption{Spectral Energy Distribution of SHBLJ~123511.0$-$140322}

}
\end{center}
\end{figure}

\begin{figure}[ht]
\begin{center}

\vbox{

\includegraphics[width=6.0cm,angle=-90]{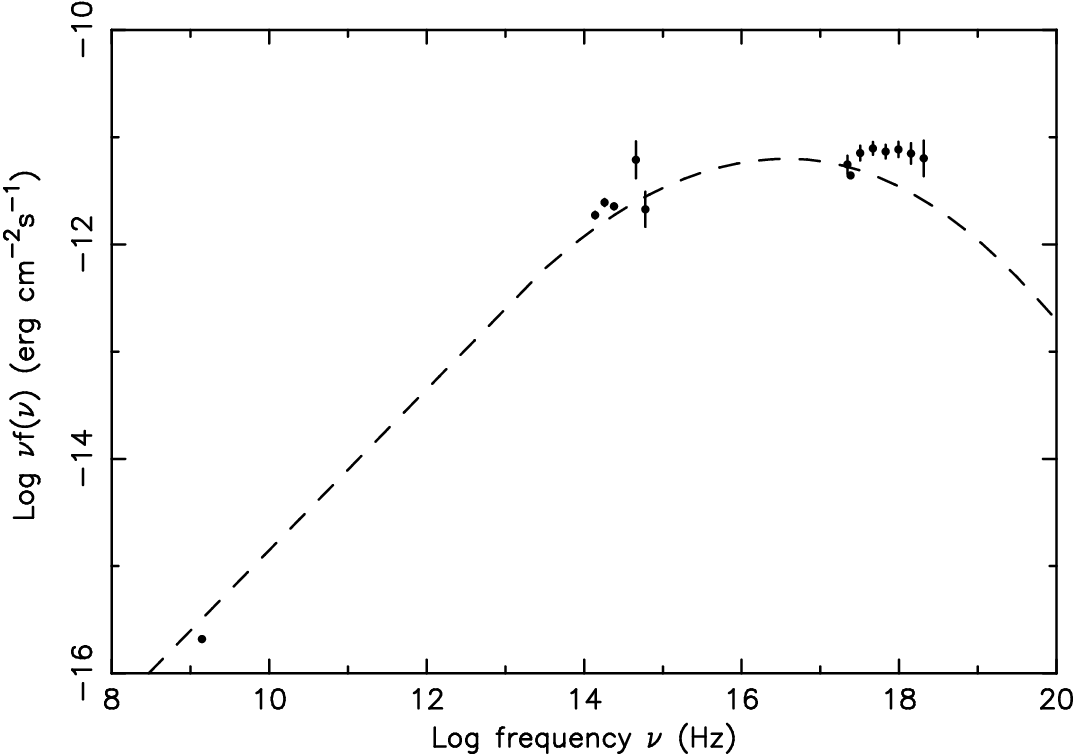}
\caption{Spectral Energy Distribution of SHBLJ~125731.9+241240}
\label{sed_1257p24}

\includegraphics[width=6.0cm,angle=-90]{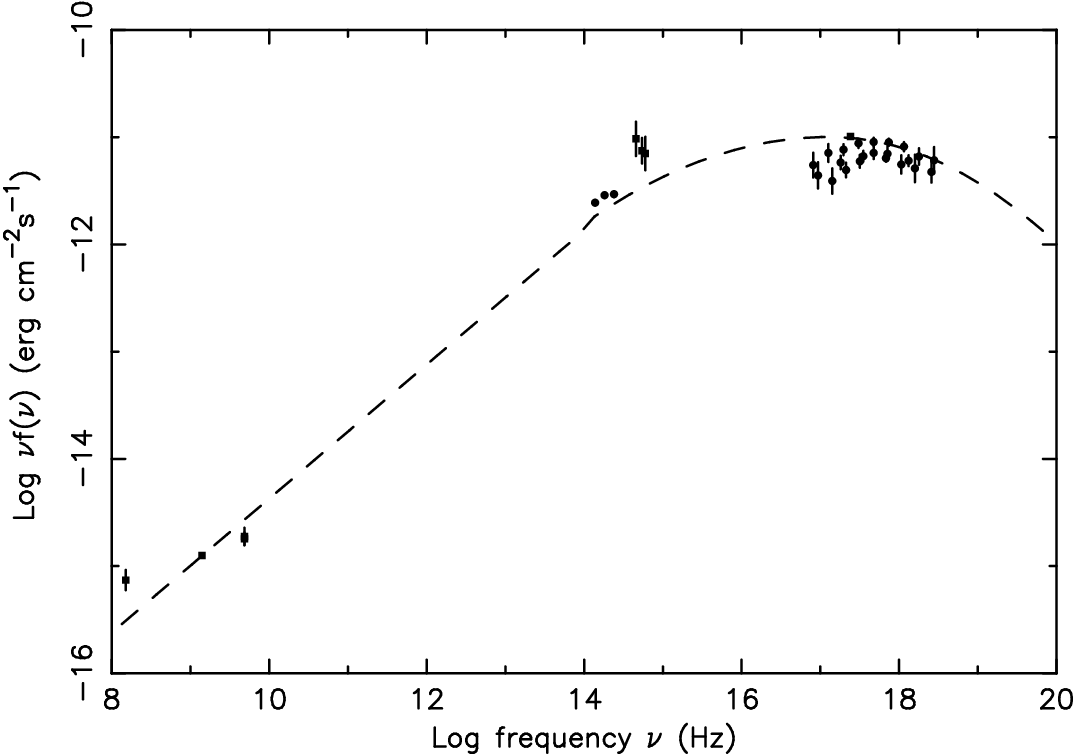}
\caption{Spectral Energy Distribution of SHBLJ~141756.1+254356}

\includegraphics[width=6.0cm,angle=-90]{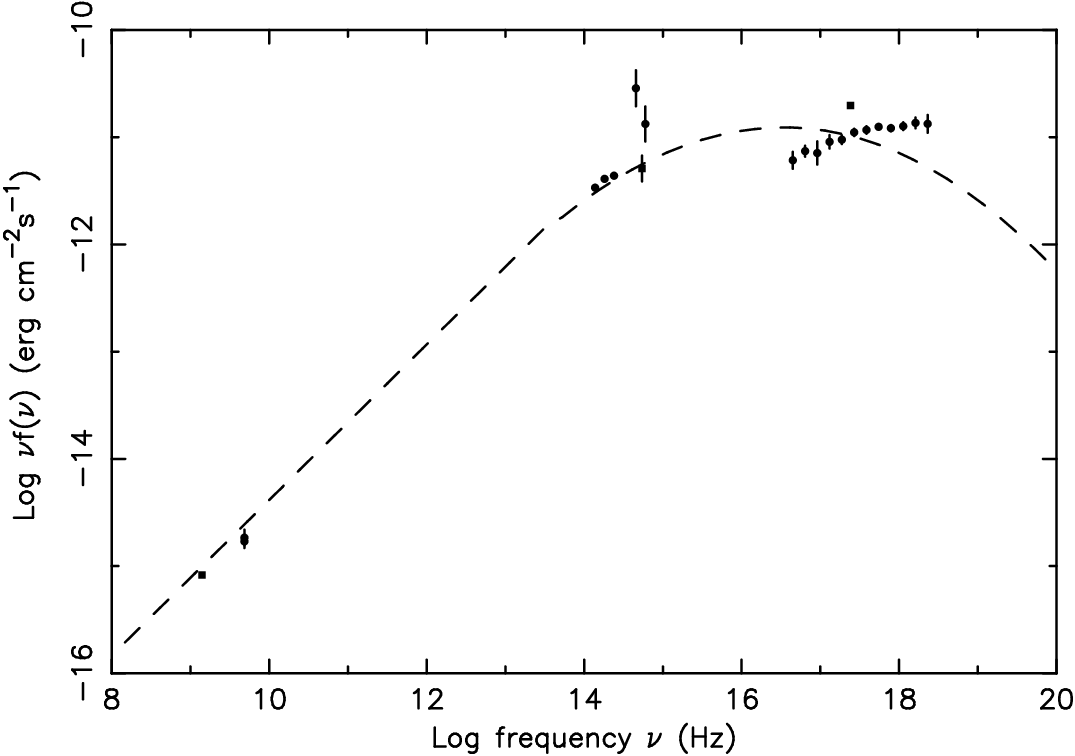}
\caption{Spectral Energy Distribution of SHBLJ~142832.6+424024}
\label{sed_1428p42}

}
\end{center}
\end{figure}

\begin{figure}[ht]
\begin{center}
\vbox{
\includegraphics[width=6.0cm,angle=-90]{sed_J1517p6525.ps}
\caption{Spectral Energy Distribution of SHBLJ~151747.4+652523}
\label{sed_1517p65}

\includegraphics[width=6.0cm,angle=-90]{sed_J1535p5320.ps}
\caption{Spectral Energy Distribution of SHBLJ~153500.9+532037}
\label{sed_1535p53}

\includegraphics[width=6.0cm,angle=-90]{sed_J2158m3013.ps}
\caption{Spectral Energy Distribution of SHBLJ~215852.0$-$301331=PKS~2155$-$304}
\label{sed_2158m30}

}
\end{center}
\end{figure}

\end{document}